\documentclass[12pt, a4paper,sort&compress]{article}

\usepackage[height=24cm,width=17cm, centering]{geometry}

\usepackage{amsfonts}
\usepackage{amsmath}
\usepackage{amssymb}
\usepackage{ascmac}
\usepackage{slashed}
\usepackage{dcolumn}
\usepackage{bm,here}
\usepackage{cite}
\usepackage{comment}
\usepackage{ifpdf}
\usepackage{slashed}
\usepackage{colortbl}
\usepackage{xcolor}
\usepackage[utf8]{inputenc}
\usepackage[mathscr]{eucal}
\usepackage[utopia]{mathdesign}

\ifpdf        
  \usepackage{graphicx}     
  \usepackage[bookmarksopen,colorlinks=true,linkcolor=dark_blue,
  citecolor=dark_red,urlcolor=dark_green,linktocpage=false]{hyperref}
\else     
  \usepackage[dvipdfmx]{graphicx}     
  \usepackage[dvipdfmx,bookmarksopen,colorlinks=true,linkcolor=dark_blue,
  citecolor=dark_red,urlcolor=dark_green,linktocpage=false]{hyperref}
\fi

\usepackage{multicol}
\definecolor{dark_blue}{rgb}{0.0, 0., 0.6}
\definecolor{dark_red}{rgb}{0.7, 0., 0.}
\definecolor{dark_green}{rgb}{0., 0.45, 0.3}
\definecolor{red}{rgb}{1,0,0}
\definecolor{blue}{rgb}{0,0,1}
\definecolor{orange}{rgb}{1,0.5,0}
\definecolor{ppink}{rgb}{1,0.4,0.4}
\definecolor{bblue}{rgb}{0.284602,0.317763,0.963947}
\usepackage{framed}
\definecolor{shadecolor}{rgb}{0.95,0.95,0.95}

\makeatletter
\@addtoreset{equation}{section}

\makeatother

\newcommand{\vev}[1]{ \left\langle {#1} \right\rangle }

\newcommand{\ket}[1]{ | {#1} \rangle }

\newcommand{\prn}[1]{\left( {#1} \right)}
\newcommand{\com}[1]{\left[ {#1} \right]}

\newcommand{\der}{\partial}  
\newcommand{\dd}{\mathrm{d}}
\newcommand{\Mpl}{M_{\rm Pl}}

\newcommand{\abs}[1]{\left\vert {#1} \right\vert}

\def\Mpl{M_{\rm Pl}}

\newcommand{\bea}{\begin{array}}
\newcommand{\eea}{\end{array}}
\newcommand{\beq}{\begin{eqnarray}}
\newcommand{\eeq}{\end{eqnarray}}

\def \psiL{\psi^{\ell}}
\def \psiLc{\psiL_{\text{c}}}
\def \psiT{\psi^t}
\def \vecpsiT{\vec{\psi}^{t}}

\def \tr{\text{Tr}}

\begin{document}

\begin{titlepage}

\begin{center}

\hfill UT 16-27\\
\hfill IPMU16-0137\\
\hfill KIAS-P16069\\

\vskip 1.in

{\LARGE \bf 
Nonthermal Gravitino Production\\[.6em]  after Large Field Inflation
}

\vskip .75in

{\large Yohei Ema$^a$, Kyohei Mukaida$^b$, Kazunori Nakayama$^{a,b}$, Takahiro Terada$^c$}

\vskip .35in

\begin{tabular}{ll}
$^{a}$&\!\! {\it Department of Physics, Faculty of Science, }\\
& {\em The University of Tokyo,  Bunkyo-ku, Tokyo 133-0033, Japan}\\[.3em]
$^{b}$ &\!\! {\it Kavli IPMU (WPI), UTIAS,}\\
&{\em The University of Tokyo,  Kashiwa, Chiba 277-8583, Japan}\\[.3em]
$^{c}$ &\!\! {\it School of Physics, Korea Institute for Advanced Study (KIAS),}\\
& {\em 85 Hoegiro, Dongdaemun-gu, Seoul 02455, Republic of Korea}\\[.3em]
\end{tabular}

\end{center}
\vskip .5in

\begin{abstract}
\noindent
We revisit the nonthermal gravitino production at the (p)reheating stage after inflation. Particular attention is paid to large field inflation models with a $\mathbb{Z}_2$ symmetry, for which the previous perturbative analysis is inapplicable; and inflation models with a stabilizer superfield, which have not been studied non-perturbatively.
It is found that in single-superfield inflation models (without the stabilizer field), nonthermal production of the transverse gravitino can be cosmologically problematic
while the abundance of the longitudinal gravitino is small enough.
In multi-superfield inflation models (with the stabilizer field), production of the transverse and longitudinal gravitinos is significantly suppressed, 
and they are cosmologically harmless.
We also clarify the relation between the background field method used in the preheating context and the standard perturbative decay method 
to estimate the gravitino abundance.
\end{abstract}


\end{titlepage}

\tableofcontents
\thispagestyle{empty}
\renewcommand{\thepage}{\arabic{page}}
\renewcommand{\thefootnote}{$\natural$\arabic{footnote}}
\setcounter{footnote}{0}

\newpage
\setcounter{page}{1}

\section{Introduction and summary}

\subsection{Introduction}

Supersymmetric (SUSY) models are well-motivated as a physics beyond the standard model,
since it provides a successful gauge coupling unification, dark matter candidate, a great reduction of the hierarchy problem etc.
In supergravity, however, there is a cosmological problem associated with the gravitino, the superpartner of the graviton,
called the gravitino problem~\cite{Pagels:1981ke, Weinberg:1982zq, Khlopov:1984pf,Ellis:1984eq}.
If the gravitino is not the lightest SUSY particle (LSP), it can decay into lighter SUSY particles.
The lifetime of the gravitino is given by~\cite{Moroi:1995fs}
\begin{align}
	\tau_{3/2} =\left(\frac{3}{8\pi} \frac{(m_{3/2}^0)^3}{\Mpl^2}\right)^{-1} \simeq 3\times 10^{-2}\,{\rm sec}\left( \frac{100\,{\rm TeV}}{m_{3/2}^0} \right)^3,
\end{align}
where $\Mpl$ is the reduced Planck scale and $m_{3/2}^0$ denotes the present gravitino mass.\footnote{
	Throughout this paper, we distinguish the ``present gravitino mass'' $m_{3/2}^0$
	and ``gravitino mass'' $m_{3/2}$, since the notion of gravitino and its mass is time-dependent in a cosmological evolution.
	The former corresponds to the gravitino mass in the present universe and it is just a constant
	while the latter is time-dependent.
}
Here we have assumed that the gravitino decays into only gaugino plus gauge boson pairs.
If other decay modes are open, the lifetime becomes slightly shorter.
Therefore, if the gravitino is much lighter than $100\,$TeV, it decays after the beginning of big-bang nucleosynthesis (BBN)
and may affect light element abundances~\cite{Moroi:1995fs,Jedamzik:2004er, Kawasaki:2004yh, Kawasaki:2004qu, Jedamzik:2006xz, Kawasaki:2008qe}.
If it is heavier, the decay itself does not affect BBN but LSPs produced by the gravitino decay can be overabundant compared with the observed dark matter abundance.
If the gravitino is LSP and $R$-parity is conserved, the gravitino itself contributes to the dark matter abundance. 
In any case, there is a strict upper bound on the gravitino abundance.

There are several processes that produce gravitinos in the early universe.
One of the unavoidable production mechanisms is thermal production:
in the high-temperature universe, scatterings of high-energy particles produce gravitinos~\cite{Bolz:2000fu,Pradler:2006qh,Rychkov:2007uq}.\footnote{
	Ref.~\cite{Ellis:2015jpg} discussed the gravitino production by the scatterings of energetic inflaton decay products 
	during the process of thermalization~\cite{Harigaya:2013vwa} and found that it is subdominant compared with the standard thermal production.
} 
The abundance of thermally produced gravitinos is proportional to the reheating temperature after inflation $T_{\rm R}$,
and hence we obtain an upper bound on $T_{\rm R}$ to avoid the gravitino problem.

Gravitinos can also be produced nonthermally.
Nonthermal gravitino production by the direct decay of inflaton was extensively studied in a series of works~\cite{Endo:2006zj,Nakamura:2006uc,Kawasaki:2006gs, Asaka:2006bv,Dine:2006ii,Endo:2006tf,Kawasaki:2006hm,Endo:2007ih,Endo:2007sz}. 
It was found that the inflaton generally decays into the gravitino pair with the partial decay rate~\cite{Kawasaki:2006gs, Asaka:2006bv, Endo:2006tf,Kawasaki:2006hm}
\begin{align}
	\Gamma(\phi\to \psi\psi) \simeq \frac{1}{64\pi}\left( \frac{\left<\phi\right>}{\Mpl} \right)^2 \frac{m_\phi^3}{\Mpl^2},  \label{inf_dec}
\end{align}
where $m_\phi$ is the inflaton mass and $\left<\phi\right>$ is the vacuum expectation value (VEV) of the inflaton.
It gives a stringent constraint on inflation models, although there are some loopholes~\cite{Endo:2007cu,Nakayama:2012hy}.
This expression for the decay rate is valid for small-field inflation models such as new inflation or hybrid inflation.\footnote{
	Chaotic inflation without a $\mathbb{Z}_2$ symmetry also leads to a similar expression.	
}

On the other hand, large field inflation models~\cite{Linde:1983gd} 
attract lots of attentions in view of recent developments on successful inflation model building 
in the framework of supergravity~\cite{Kawasaki:2000yn,Kallosh:2010ug,Kallosh:2010xz,Ferrara:2010in,Nakayama:2010kt,Nakayama:2013txa,Kallosh:2013hoa,Kallosh:2013yoa,Galante:2014ifa}.
It is interesting because it can be tested with on-going/future $B$-mode polarization experiments.
In large field inflation models with a $\mathbb{Z}_2$ symmetry in which the inflaton field oscillates around the origin $\phi=0$ after inflation,
we cannot use the expression (\ref{inf_dec}) as a gravitino production rate.
This is simply because the calculations in Refs.~\cite{Endo:2006tf,Kawasaki:2006hm,Endo:2007ih,Endo:2007sz} assumed
the perturbative decay of inflaton around its VEV.
In inflation models with the $\mathbb{Z}_2$ symmetry, however, there is no such decay process due to the $\mathbb{Z}_2$ symmetry.\footnote{Another assumption there was that the SUSY is dominantly broken by the Polonyi field
at the end of reheating so that the definition of ``gravitino'' at that epoch is the same as the present-day gravitino.
This assumption is valid as long as we are interested in the gravitino production at the end of reheating.
In large field inflation models, however, the gravitino production is often dominated at the epoch just after inflation (preheating)
and hence this assumption is not justified, as we will see later.}

This does not mean that the inflaton cannot decay into gravitinos as well as other light particles.
The inflaton coherent oscillation affects the masses or kinetic terms of coupled particles.
The coupled particles, including gravitinos, ``feel'' the rapid inflaton oscillation and it affects the evolution 
of their wave functions.
It is known that this leads to particle production, often in the context of preheating~\cite{Dolgov:1989us,Traschen:1990sw,Shtanov:1994ce,Kofman:1994rk,Kofman:1997yn}.
Therefore, even if the inflaton has the $\mathbb{Z}_2$ symmetry, its coherent oscillation necessarily transfers its energy to the coupled particles.
The question we would like to address is: 
what amount of gravitinos is produced during the preheating?

Production of gravitinos during the preheating was first discussed 
in Refs.~\cite{Kallosh:1999jj,Giudice:1999am} in a single-superfield case, 
in which only one inflaton chiral superfield was introduced.
There it was found that in the preheating stage, longitudinal gravitinos are efficiently produced.
Later it was recognized that the theory of gravitino preheating is much more involved due to the subtlety of the notion of ``gravitino''~\cite{Kallosh:2000ve,Nilles:2001ry,Nilles:2001fg}.
The gravitino becomes massive by ``absorbing'' the goldstino, but the definition of goldstino is time-dependent in a cosmological background.
In the early universe, the inflaton oscillation energy dominantly breaks SUSY and hence the goldstino is almost the inflatino, 
the fermionic superpartner of the inflaton. 
At late time, the Polonyi field\footnote{
	In this paper we call the present-day SUSY breaking field as Polonyi field.
} dominantly breaks SUSY
and its fermionic component, Polonyino, becomes the goldstino.
Thus the composition of goldstino changes with time.
Refs.~\cite{Nilles:2001ry,Nilles:2001fg} noticed that it is essential to include (at least) two chiral superfields, inflaton and Polonyi, 
to correctly deal with this problem and concluded that 
what the preheating efficiently produces eventually becomes the inflatino, 
which is less harmful than the gravitino.

Still, however, a quantitative/comprehensive analysis of the nonthermal gravitino production rate in such a case is missing.
Although Refs.~\cite{Nilles:2001ry,Nilles:2001fg} revealed that the gravitino production is suppressed than previously thought,
it is highly non-trivial how we can extrapolate their numerical results into more realistic setups 
and parameters both in the inflaton and SUSY breaking sector.
Thus we would like to 
provide general analytic formulae for the nonthermal gravitino abundance that are applicable to any realistic models.

\subsection{Brief summary}

In this paper, we revisit the theory of nonthermal gravitino production in a comprehensive and unified manner.
Our purposes and results are summarized below.
\begin{itemize}
\item
We derive nonthermal gravitino production rates and their resulting abundances quantitatively with useful formulae.
We find that in single-superfield inflation models, the production of transverse gravitino is significant and cosmologically problematic,
while the production of longitudinal gravitino is less important.
This aspect of the nonthermal gravitino production has been overlooked in previous literatures except for a few~\cite{Maroto:1999ch}.
\item
Recent realistic large field inflation models introduce an additional chiral superfield, called a stabilizer~\cite{Kawasaki:2000yn,Kallosh:2010ug,Kallosh:2010xz,Ferrara:2010in,Nakayama:2010kt,Nakayama:2013txa,Kallosh:2013hoa,Kallosh:2013yoa,Galante:2014ifa}.
We find that in models with the stabilizer field, the production of transverse gravitino is significantly suppressed and it is cosmologically harmless.
For the longitudinal component, the production rate is similar to the single-superfield case.
\item
We show the equivalence between the background field method developed in Refs.~\cite{Kallosh:2000ve,Nilles:2001ry,Nilles:2001fg}
and the perturbative decay method developed in Refs.~\cite{Endo:2006tf,Kawasaki:2006hm,Endo:2007ih,Endo:2007sz} for evaluating the gravitino abundance in some sense.
The former can deal with a broad class of models including $\mathbb{Z}_2$-symmetric large field models, while in models without the $\mathbb{Z}_2$ symmetry, it gives the same result as the perturbative decay method.
\end{itemize}

This paper is organized as follows.
In Sec.~\ref{sec:grav}, we review the structure of the gravitino Lagrangian to set the stage of discussion in the subsequent Sections.
We formulate a general setup to discuss multi-superfield case.
Gravitino production in the single-superfield inflation models and in the multi-superfield inflation models are studied in Sec.~\ref{sec:single} and Sec.~\ref{sec:multi}, respectively.
The analyses include both the transverse and longitudinal components of gravitino.
Our conclusion is in Sec.~\ref{sec:conclusion}, and the gravitino abundance is summarized in Fig.~\ref{fig:Y}.
App.~\ref{sec:notation} summarizes our notations and conventions.
The background field method to evaluate the fermion production rate, often used in the preheating context, is reviewed in App.~\ref{sec:fermion}.
We also briefly cover the gravitino production in small-field inflation models in App.~\ref{sec:small-field}
in order to show the equivalence between our method and the perturbative decay method.
In App.~\ref{sec:dyn}, we review the multi-field scalar dynamics to discuss the induced oscillation of the Polonyi field in the main text.
Calculations of mass eigenvalues are given in App.~\ref{sec:mass}, which are used in Secs.~\ref{sec:long_single} and \ref{sec:long_multi}.

\section{Gravitino Lagrangian}  \label{sec:grav}

\subsection{Master supergravity Lagrangian} \label{sec:master}
We start from the following supergravity Lagrangian:
\begin{align}
	e^{-1}\mathcal{L} 
	=& 
	\frac{\Mpl^2}{2}R - g_{i\bar{j}}\partial_{\mu}\phi^{i} \partial^{\mu}\phi^{*}{}^{\bar{j}} - V \nonumber \\
	&-\frac{1}{2}\overline{\psi}_\mu R^{\mu} + \frac{1}{2}\overline{\psi}_\mu \left(m_{3/2} P_R + m_{3/2}^* P_L\right)
	\widehat{\gamma}^{\mu\nu}\psi_\nu \nonumber \\
	&-\frac{1}{2}g_{i\bar{j}}\left(\overline{\chi}^{i}_{L}\widehat{\slashed{D}}\chi^{\bar{j}}_{R} + \overline{\chi}^{\bar{j}}_{R}\widehat{\slashed{D}}\chi^{i}_{L}\right)
	-\frac{1}{2}\left(m_{ij}\overline{\chi}^{i}_{L}\chi^{j}_{L} + m_{\bar{i}\bar{j}}\overline{\chi}^{\bar{i}}_{R}\chi^{\bar{j}}_{R}\right) \nonumber \\
	&+ \frac{\sqrt{2}}{\Mpl}g_{i\bar{j}}\overline{\psi}_\mu \widehat{\gamma}^{\nu\mu}
	\left(\chi^{i}_{L}\partial_{\nu}\phi^{*\bar{j}} + \chi^{\bar{j}}_{R}\partial_\nu \phi^{i}\right)
	+ \frac{1}{\sqrt{2}\Mpl}\left(\overline{\psi}\cdot \widehat{\gamma}\right) v + e^{-1}\mathcal{L}_{4f},
	\label{eq:master_lagrangian}
\end{align}
where $e$ is the determinant of the vierbein $e_{\mu}^a$, 
$\Mpl$ is the reduced Planck mass, $R$ is the Ricci scalar,  $\phi^{i}$ and $\phi^{*\bar{i}}$ are scalar fields and their complex conjugates, $\chi^{i}_{L}$ and $\chi^{\bar{j}}_{R}$ are left-handed matter fermions and their conjugate right-handed fermions, and $g_{i\bar{j}}=\partial_i \partial_{\bar{j}} K$ is the K\"ahler metric with $\partial_i$ being the derivative with respect to $\phi^i$.
The scalar potential $V$ is given in terms of the K\"ahler potential $K$ and the superpotential $W$ as
\begin{align}
	V = e^{K/\Mpl^2}\left[g^{i\bar{j}}D_i W D_{\bar{j}} W^{*} - \frac{3\left\lvert W \right\rvert^2}{\Mpl^2}\right],
\end{align}
where $D_i W = \partial_i W + \left(\partial_i K\right)W/\Mpl^2$,
 and $g^{i\bar{j}}$ is the inverse K\"ahler metric.
The hats denote the quantities in the curved space-time.
The field strength of the gravitino $\psi_\mu$ in its kinetic term is given by
\begin{align}
	R^\mu \equiv& \widehat{\gamma}^{\mu\rho\sigma}D_\rho \psi_\sigma, \\
	D_\mu \psi_\nu =& \left(\partial_{\mu} + \frac{1}{4}\omega_{\mu}{}^{ab}\gamma_{ab} 
	- i A_\mu \gamma_* \right)\psi_\nu   
	- \Gamma_{\mu\nu}^{\rho}\psi_\rho,
\end{align}
where
\begin{align}
	A_\mu \equiv \frac{i}{4\Mpl^2}\left( \partial_i K \partial_\mu \phi^i - \partial_{\bar{i}} K \partial_\mu \phi^{*\bar{i}}\right),
\end{align}
is the ``remnant'' of the gauge field of the $R$-symmetry in the underlying superconformal formulation,
and $\gamma_a$ is the Dirac gamma matrix in the flat space.
Note that practically the Christoffel symbol does not contribute due to the anti-symmetry of $\widehat{\gamma}^{\mu\rho\sigma}$.
The gravitino mass $m_{3/2}$ and the other fermion mass matrix $m_{ij}$ are respectively defined as
\begin{align}
	m_{3/2} &\equiv e^{K/2\Mpl^2}\frac{W}{\Mpl^2}, \\
	m_{ij} &\equiv e^{K/2\Mpl^2}\left[ \partial_i + \frac{\partial_i K}{\Mpl^2}\right]D_j W - e^{K/2\Mpl^2}\Gamma_{ij}^{k}D_k W,
\end{align}
where $\Gamma_{ij}^{k} \equiv g^{k\bar{l}} \partial_{i}g_{j\bar{l}}$ is the Christoffel symbol in the K\"{a}hler manifold.
The goldstino $v$ is defined as
\begin{align}
	v_L \equiv e^{K/2\Mpl^2} D_i W\chi^{i}_{L} + g_{i\bar{j}}\slashed{\partial}\phi^i \chi^{\bar{j}}_{R}.
\end{align}
The last term in Eq.~\eqref{eq:master_lagrangian}, $\mathcal{L}_{4f}$, denotes the four-fermion interactions originating from the torsion, which we will neglect from now.
We consider only gauge singlet components in this paper. For more details on the notation and conventions used in this paper,
see App.~\ref{sec:notation}.

The master Lagrangian~\eqref{eq:master_lagrangian} contains the gauge redundancy as well as 
unphysical degrees of freedom. From now, we fix the fermionic gauge redundancy by taking the unitary gauge
\begin{align}
	v = 0,
\end{align}
and integrate out the unphysical degrees of freedom to obtain the physical Lagrangian.
In order to do so, we should first find constraint equations, and solve them to express 
the unphysical degrees of freedom in terms of the physical degrees of freedom, \textit{i.e.,} the transverse
and longitudinal modes of the gravitino. Below we give the outline of this procedure. 
For more details, see Ref.~\cite{Kallosh:2000ve}.
In the following, we assume that the scalar fields are real for simplicity, and hence 
$m_{3/2}^{*} = m_{3/2}$ and $A_\mu = 0$.

\subsubsection*{Constraint equations}
In the unitary gauge, the equations of motion for the gravitino are given by
\begin{align}
	\Sigma^{\mu} \equiv R^\mu - \widehat{\gamma}^{\mu\nu}\left(m_{3/2}\psi_\nu 
	- \frac{\sqrt{2}}{\Mpl}g_{i\bar{j}}
	\left(\chi^{i}_{L}\partial_\nu \phi^{*\bar{j}} + \chi^{\bar{j}}_{R}\partial_\nu \phi^{i}\right)\right) = 0.
\end{align}
From these equations, we can verify that the following equations do not contain the time derivatives
with respect to the gravitino, and hence are constraints:
\begin{align}
	0 &= D_\mu \Sigma^\mu + \frac{m_{3/2}}{2}\widehat{\gamma}_\mu \Sigma^\mu, \label{eq:constraint1} \\
	0 &= \Sigma^{0}. \label{eq:constraint2}
\end{align}
The first constraint~\eqref{eq:constraint1} can be solved to give $\psi_0$ in terms of $\vec{\psi}$,
$\chi^{i}_{L}$ and $\chi^{\bar{j}}_{R}$. Practically, however, we do not need to know the explicit solution for $\psi_0$.
This is because the Lagrangian~\eqref{eq:master_lagrangian} without $\mathcal{L}_{4f}$ 
depends only linearly on $\psi_0$, and hence $\psi_0$ contributes to the Lagrangian 
with a combination of $\psi_0 \Sigma^{0}$.
Thus, it automatically drops from the Lagrangian once we impose the second constraint~\eqref{eq:constraint2}.
For this reason, we concentrate only on the second constraint~\eqref{eq:constraint2}, which relates $\vec{k}\cdot \vec \psi$ and $\vec \gamma \cdot \vec \psi$.

So far we took the background metric to be generic. From now on, we take it to be the 
Friedmann-Lema\^itre-Robertson-Walker (FLRW) one,
\begin{align}
	\text{d}s^2 = -\text{d}t^2 + a^2(t)\text{d}\vec{x}^2 = a^2(\eta)( -\text{d}\eta^2 + \text{d}\vec{x}^2),
\end{align}
with $a$ being the scale factor,
since we are interested in the gravitino production in the cosmological background.
We also decompose the gravitino as
\begin{align}
	\vec{\psi} = \vecpsiT + \left(\frac{1}{2}\vec{\gamma} 
	- \frac{1}{2k^2}\vec{k}\left(\vec{k}\cdot\vec{\gamma}\right)\right)\psiL
	+ \left(\frac{3}{2k^2}\vec{k} - \frac{1}{2k^2}\vec{\gamma}\left(\vec{k}\cdot\vec{\gamma}\right)\right)
	\vec{k}\cdot\vec{\psi},
\end{align}
where the longitudinal mode is $\psiL \equiv \vec{\gamma}\cdot \vec{\psi}$
and the transverse mode satisfies $\vec{\gamma}\cdot\vecpsiT = \vec{k}\cdot\vecpsiT = 0$, respectively.
Here we have moved to the momentum space.
Then, we can solve the second constraint~\eqref{eq:constraint2}
to obtain $\vec{k}\cdot\vec{\psi}$ in terms of $\psiL$ as
\begin{align}
	i\vec{k}\cdot\vec{\psi} = \left(i\vec{\gamma}\cdot\vec{k} - a\left(m_{3/2} + \gamma_0 H\right)\right)\psiL,
\end{align}
where $H \equiv \dot{a}/a$ is the Hubble parameter.\footnote{
	In this paper, the dot denotes the derivative with respect to time $t$ while $\partial_0$ represents the derivative 
	with respect to conformal time $\eta$.
}
This relation ensures that $\psi^\ell$ is actually ``longitudinal''.
By substituting it to the original Lagrangian and sorting things out,
we obtain the Lagrangian for the physical gravitino $\vecpsiT$
and $\psiL$. Its explicit form will be shown in the next subsection.

\subsection{Lagrangian of physical gravitino}

The gravitino Lagrangian in terms of the physical degrees of freedom is given by
\begin{align}
	e^{-1}\mathcal{L}_{3/2}
	=&
	e^{-1}\mathcal{L}_{t} + e^{-1}\mathcal{L}_{\ell} + e^{-1}\mathcal L_{\rm mix},
\end{align}
where the first two terms correspond to the kinetic and mass terms of the transverse and longitudinal gravitino, respectively:\footnote{
	Here we keep terms which are exactly zero due to the Majorana property (\textit{e.g.}, $\bar \psi \gamma^0 \psi = 0$),
	in order for the combinations which appear in the equations of motion to be manifest.
}
\begin{align}
	e^{-1}\mathcal{L}_{t}
	=&
	- \frac{1}{2a^3}\overline{\vecpsiT}\slashed{D}\vecpsiT 
	+\frac{H}{2a^2} \overline{\vecpsiT}\gamma^{0}\vecpsiT
	-\frac{1}{2a^2}\overline{\vecpsiT}m_{3/2}\vecpsiT,
\end{align}
and
\begin{align}
	e^{-1}\mathcal{L}_{\ell}
	=&
	-\frac{\rho_\mathrm{SB}}{4ak^2\Mpl^2}\overline{\psiL}\left[
	\gamma^{0}\partial_{0} + \left(i\vec{\gamma}\cdot\vec{k}\right)\widehat{A}
	-\frac{3a}{2}\left(m_{3/2}+ H\gamma^{0}\right)\widehat{A} 
	- \frac{1}{2}a m_{3/2} 
	\right]\psiL.
\end{align}
The last term represents the mixing between the longitudinal gravitino and chiral fermions:
\begin{align}
	e^{-1}\mathcal L_{\rm mix} = \frac{\sqrt{2}}{a^2\Mpl}\overline{\psiL}\gamma^{0} g_{i\bar j}
	\left[(\partial_0\phi^i)\chi^{{\bar j}}_{R}+(\partial_0 \phi^{*\bar j})\chi^{i}_{L} \right].
\end{align}
Note that the transverse mode does not mix with chiral fermions.
Here we have defined
\begin{align}
	&\widehat A \equiv \frac{p_{\rm SB} - \gamma^0 p_{W}}{\rho_{\rm SB}},  \label{Ahat} \\
	&\rho_{\rm SB} \equiv \sum_i|\dot\phi_i|^2 + V_{\rm SB},~~~V_{\rm SB} \equiv V + 3m_{3/2}^2 \Mpl^2 = \sum_i |F_i|^2,\\
	&p_{\rm SB} \equiv \sum_i|\dot\phi_i|^2 - V_{\rm SB},~~~p_{W} \equiv 2\dot m_{3/2} \Mpl^2 =-\sum_i(\dot\phi_i^* F_i + \dot \phi_i F_i^*),
\end{align}
where the SUSY breaking $F$-term is defined as $F^i = - e^{K/2\Mpl^2} g^{i\bar{j}} D_{\bar{j}}W^*$, 
for the minimal K\"ahler potential.\footnote{
	In this paper the ``minimal'' K\"ahler potential means $K_{i\bar j} \simeq \delta_{i\bar j}$.
	Adding a holomorphic function in $K$ does not change the results.
	Note that the minimal shift-symmetric K\"{a}hler potential ($K=-(\phi - \phi^{\dag})^2/2$) also satisfies $K_{i\bar{j}}=\delta_{i\bar{j}}$.
	Also we will introduce a higher order K\"ahler potential like $K \sim |z|^4/\Lambda^2$ for a Polonyi field $z$,
	but it does not affect the discussion as long as the field value of $z$ is small enough.
} 
 The $\widehat{A}$ is an important combination whose phase rotation gives rise to gravitino production as we will see below.
The $V_{\text{SB}}$, $\rho_{\text{SB}}$ and $p_{\text{SB}}$ can be interpreted as the potential energy, energy density and pressure of the matter components which break SUSY while $p_W$ measures the time variation of the gravitino mass, which can also be expressed like the ``geometric average'' of the SUSY breaking kinetic and potential energies.
Here and in what follows, we assume the reality of dynamical scalar fields for simplicity (and accordingly, the difference between the subscript and superscript of the field index disappears).
Actually, if the parameters in the K\"ahler and superpotentials are all real and the initial condition of the fields is taken to be real,
the subsequent dynamics does not affect the reality of scalar fields.
The Friedmann equation reads
\begin{align}
	3H^2 \Mpl^2 = \rho = \sum_i|\dot\phi_i|^2 + V.
\end{align}
Here $\rho$ is the energy density of the system. 
Note that $\rho_{\rm SB}$ is the energy density associated with SUSY breaking, which differs from $\rho$.
It also means that
\begin{align}
	3(H^2+m_{3/2}^2)\Mpl^2 = \rho_{\rm SB}.
\end{align}

Now let us rewrite these Lagrangians in terms of the canonical field for later convenience.
The Lagrangian of the transverse mode in terms of the canonical field $\vecpsiT_{\text{c}} \equiv \sqrt{a}\vecpsiT$ is given by\footnote{
	Note that $\slashed{D}\psi = (\slashed{\partial} + \frac{3}{2}aH\gamma^0)\psi$ and $\bar\psi \gamma^0 \psi = 0$ for a Majorana fermion $\psi$.
}
\begin{align}
	\mathcal L_t = -\frac{1}{2}\overline{\vecpsiT_{\text{c}}} \left[ \gamma^0 \partial_0 + i \vec\gamma\cdot\vec k + a m_{3/2} \right] \vecpsiT_{\text{c}}.
	\label{Ltrans}
\end{align}
Note that the mass of the order of the Hubble scale disappears reflecting the conformal invariance.
Thus it does not ``feel'' the Hubble expansion, meaning that the oscillation/non-adiabatic change of the Hubble parameter 
does not lead to the production of transverse gravitino.
On the other hand, the gravitino mass $m_{3/2}$ oscillates rapidly in the inflaton oscillation epoch, which inevitably leads to significant gravitino production.

As for the longitudinal mode, let us rewrite the Lagrangian by using the canonical field
\begin{align}
	\psiLc \equiv - \frac{\sqrt{\rho_{\rm SB}} a^{3/2}}{\sqrt{2}k^2\Mpl} i\left( \vec\gamma\cdot\vec k\right)\psiL,
\end{align}
as
\begin{align}
	\mathcal L_\ell = - \frac{1}{2}\overline{\psiLc} \left[ 
		\gamma^0\partial_0 - i\left( \vec\gamma\cdot\vec k\right)\widehat A^{\dag} 
		+ a\widehat{m}_{3/2}
	\right]\psiLc,  
	\label{grav_kin}
\end{align}
and
\begin{align}
	\mathcal L_{\rm mix} =& \frac{2a^{1/2}}{\sqrt{\rho_{\rm SB}}}\overline{\psiLc} i\left(\vec \gamma\cdot\vec k \right)\gamma^0
	g_{i\bar j}\left[(\partial_0\phi^i)\chi^{{\bar j}}_{R}+(\partial_0 \phi^{*\bar j})\chi^{i}_{L} \right] \nonumber \\
	=& \frac{2}{ \sqrt{\rho_{\rm SB}}}\overline{\psiLc} i\left(\vec \gamma\cdot\vec k \right)\gamma^0
	g_{i\bar j}\left[\dot{\phi}^i \chi'{}^{{\bar j}}_{R}+\dot{\phi}^{*\bar j} \chi'{}^{i}_{L} \right].
	\label{Lmix}
\end{align}
In the second equality, chiral fermions in the original supergravity Lagrangian $\chi^i$ are rescaled 
as $\chi'{}^i \equiv a^{3/2} \chi^i$ and the prime is dropped in the following discussion so that their kinetic terms become 
$\mathcal L_f = -g_{i\bar j}(\overline\chi_L^{\bar j} \slashed{\partial}\chi_L^i +\overline\chi_R^{i} \slashed{\partial}\chi_R^{\bar j})/2$ without explicit dependence on the scale factor $a$.
In Eq.~\eqref{grav_kin}, we have defined the generalized gravitino mass term,
\begin{align}
	\widehat{m}_{3/2} \equiv 
	\frac{3Hp_{W} + m_{3/2}(\rho_{\rm SB}+3p_{\rm SB})}{2\rho_{\rm SB}}
	\label{hatm3/2},
\end{align}
which is of the order of $\widehat{m}_{3/2} \sim \mathcal O(H) + \mathcal O(m_{3/2})$.
In the Lagrangian, the coefficient $\widehat A$ as well as the gravitino mass $m_{3/2}$ oscillates rapidly.
The whole structure is slightly complicated, especially in the multi-superfield case.
However, in almost all the situations of our interest, the SUSY breaking is dominated by one field 
and the problem effectively reduces to the single-superfield case, in which the analysis is significantly simplified.

\subsection{The case of only one chiral superfield}  \label{sec:singlegrav}

In order to illustrate the structure of the theory, first let us focus on the single-superfield case. 
In this case, we have a relation
\begin{align}
	|\widehat A|^2 = \frac{p_{\rm SB}^2 + p_{W}^2}{\rho_{\rm SB}^2}= 1.  \label{A^2=1}
\end{align}
Therefore we can take $-\widehat A \equiv e^{2\gamma^0 \theta}$ with $\theta$ being a real parameter.
Then by defining $\psiLc{}' \equiv e^{-\gamma^0\theta} \psiLc$ (and dropping the prime thereafter) the Lagrangian is simplified as
\begin{align}
	\mathcal L_\ell = -\frac{1}{2}\overline{\psiLc} \left[ 
		\gamma^0\partial_0  +  i\vec\gamma\cdot\vec k - \partial_0\theta
		+a\widehat{m}_{3/2}
	\right]\psiLc.
\end{align}
The remaining task is to calculate $\partial_0\theta = a\dot\theta$.\footnote{
	 In the single-superfield case, there is no mixing term $\mathcal L_{\rm mix}$ because 
	 there is only one chiral fermion $\chi$ except for the gravitino, and it must be proportional to the goldstino $v$ that vanishes $(v\,=0)$ in the unitary gauge.
}
By using the condition (\ref{A^2=1}), we obtain
\begin{align}
	\dot\theta= \frac{1}{2}\frac{\rho_{\rm SB}\dot p_{\rm SB} - \dot \rho_{\rm SB} p_{\rm SB}}{ p_W \rho_{\rm SB}}.
	\label{theta_single}
\end{align}
From the energy conservation conditions etc., we obtain\footnote{
	Useful equations are: $\dot\rho=-6H|\dot\phi|^2$, $\dot\rho_{\rm SB} = \dot\rho + 3m_{3/2}p_{W}$, $\dot p_{\rm SB} = \dot\rho - 2\dot V - 3m_{3/2}p_{W}$.
}
\begin{align}
	\dot\theta = -\frac{\dot V}{p_W}
	- \frac{3m_{3/2}|\dot\phi|^2}{\rho_{\rm SB}}
	- \frac{6H|\dot\phi|^2 V_{\rm SB}}{p_W\rho_{\rm SB}}.
	\label{theta_single2}
\end{align}
Substituting the explicit expression of $\dot \theta$, we obtain
\begin{align}
	\mathcal L_\ell = -\frac{1}{2}\overline{\psiLc} \left[ 
		\gamma^0\partial_0  +  i\vec\gamma\cdot\vec k
		+\frac{a\dot V}{p_{W}}+\frac{3aH p_{W}}{\rho_{\rm SB}}
		+am_{3/2}\left(2+\frac{3p_{\rm SB}}{\rho_{\rm SB}} \right)
	\right]\psiLc.
\end{align}
Below we estimate behavior of the mass term in typical one-field dominated cases.

\subsubsection{Inflaton-dominated SUSY breaking}  \label{sec:inflaton}

Let us consider the case where the inflaton $\phi$ dominates the SUSY breaking.
As we will see later, if the energy density of the universe is dominated by the inflaton oscillation,
and $H \gg m_{3/2}^0$, the SUSY is dominantly broken by the inflaton superfield.
The K\"ahler potential and superpotential are assumed to be 
\begin{align}
	&K = |\phi|^2,\\
	&W = \frac{1}{2}m_\phi\phi^2.
\end{align}
Up to correction with $\mathcal O(\phi^2/\Mpl^2)$, we obtain
\begin{align}
	&V_\phi \simeq m_\phi(m_\phi + 5m_{3/2}) \phi,\\
	&p_{W} \simeq 2(m_\phi + 2m_{3/2}) \phi\dot\phi,
\end{align}
where $m_{3/2} \simeq m_\phi\phi^2/(2\Mpl^2)$.
Note that we have assumed the reality of $\phi$-dynamics to derive those expressions.
Substituting these equations into $\mathcal L_\ell$, we obtain\footnote{
	Note that $\dot V = V_\phi\dot\phi + V_{\phi^*}\dot\phi^* = 2V_\phi\dot\phi$ for real $\phi$.
	Here $V_\phi$ should be regarded as $\partial_\phi V(\phi,\phi^*)$.
}
\begin{align}
	\mathcal L_\ell = -\frac{1}{2}\overline{\psiLc} \left[ 
		\gamma^0\partial_0  +  i\vec\gamma\cdot\vec k + am_\phi + \frac{3aH p_{W}}{\rho_{\rm SB}}
		+ am_{3/2}\left(5+\frac{3p_{\rm SB}}{\rho_{\rm SB}} \right)
	\right]\psiLc.  \label{Ll_single_quad}
\end{align}
It obtains a mass of the order of the inflatino mass, as expected.\footnote{
	This expression is consistent with Eq.\,(9.15) of~\cite{Kallosh:2000ve}.
}
Note that in the inflaton-dominated SUSY breaking case, we roughly have $\rho_{\rm SB} \sim |p_{\rm SB}| \sim |p_{W}|$.
Especially $p_{\rm SB}$ and $p_{W}$ are violently oscillating functions with frequency of twice the inflaton oscillation,
and $m_{3/2}$ itself is also an oscillating function.
Thus, both terms proportional to $H$ and $m_{3/2}$ contribute to the production of longitudinal gravitino.

For the superpotential with a higher power,
\begin{align}
	W = \frac{1}{n}\lambda\phi^n,
\end{align}
with $n>2$, we obtain
\begin{align}
	&V_\phi \simeq (n-1)\lambda^2\phi^{2n-3}\left(1+\frac{n+3}{n}\frac{\phi^2}{\Mpl^2} \right),\\
	&p_{W} \simeq 2\lambda\phi^{n-1}\dot\phi\left(1+\frac{n+2}{2n}\frac{\phi^2}{\Mpl^2} \right),
\end{align}
up to corrections with $\mathcal O(\phi^2/\Mpl^2)$. 
The Lagrangian of the longitudinal gravitino becomes
\begin{align}
	\mathcal L_\ell = -\frac{1}{2}\overline{\psiLc} \left[ 
		\gamma^0\partial_0  +  i\vec\gamma\cdot\vec k + a(n-1)\lambda\phi^{n-2}+ \frac{3aH p_{W}}{\rho_{\rm SB}}
		+ am_{3/2}\left(\frac{n(n+3)}{2}+\frac{3p_{\rm SB}}{\rho_{\rm SB}} \right)
	\right]\psiLc, \label{Ll_single_n}
\end{align}
where $m_{3/2} \simeq \lambda\phi^n/(n\Mpl^2)$.
Note that $n=2$ reproduces the above result~\eqref{Ll_single_quad}.
In this case with general $n$, the longitudinal gravitino obtains a mass of $\sim\lambda\phi^{n-2}$, which is rapidly oscillating.
Other terms proportional to $H$ and $m_{3/2}$ are suppressed by $\phi/\Mpl$ and $\phi^2/\Mpl^2$, respectively.
Therefore, the mass term of $\sim\lambda\phi^{n-2}$ contributes to the gravitino production dominantly
and it is much more efficient than the case of $n=2$.

When we start from the minimal shift-symmetric K\"{a}hler potential, $K \supset - (\phi - \phi^\dag)^2 / 2$, the qualitative discussion is maintained.  The only differences in the final expressions of the Lagrangian are that $5$ and $n(n+3)/2$ in the last parenthesis in Eqs.~\eqref{Ll_single_quad} and \eqref{Ll_single_n} respectively are both replaced with $-1$.

\subsubsection{Polonyi-dominated SUSY breaking}  \label{sec:Polonyi}

Now let us consider the case that the Polonyi field $z$ dominates the SUSY breaking as in the present universe.
In a cosmological setup, this approximation is valid at $H\ll m_{3/2}^0$.
The K\"ahler potential and superpotential are assumed to be 
\begin{align}
	&K = |z|^2 - \frac{|z|^4}{\Lambda^2},\\
	&W = \mu^2 z + W_0.
\end{align}
In the present paper, we always implicitly assume a dynamical SUSY breaking scenario~\cite{Affleck:1983mk,Affleck:1984xz,Izawa:1996pk,Intriligator:1996pu,Intriligator:2006dd} in which the Polonyi field $z$ obtains 
a large SUSY breaking mass, represented by the nonminimal K\"ahler potential.\footnote{
	Otherwise, the cosmological Polonyi problem is much more serious than the gravitino overproduction problem~\cite{Coughlan:1983ci, Banks:1993en, deCarlos:1993wie}. Also we assume that dynamical SUSY breaking already occurred during inflation for simplicity.
}
In this model we obtain
\begin{align}
	&V_z \simeq m_z^2 z - 2\sqrt{3}(m_{3/2}^0)^2 \Mpl = m_z^2 \delta z,\\
	&p_{W} \simeq 2\mu^2 \dot z = 2\sqrt{3} m_{3/2}^0 \Mpl \dot z ,
\end{align}
where the mass of the Polonyi field is given by $m_z^2 = 12(m_{3/2}^{0} \Mpl/\Lambda)^2$ $(\gg (m_{3/2}^0)^2)$
and its VEV is $\langle z\rangle \simeq2\sqrt{3}\Mpl(m_{3/2}^0/m_z)^2$ and $\delta z \equiv z-\langle z\rangle$ with $m_{3/2}^0 \simeq \mu^2/\sqrt{3}\Mpl \simeq W_0 / \Mpl^2$.
Substituting these equations into $\mathcal L_\ell$, we obtain a simple expression:
\begin{align}
	\mathcal L_\ell = -\frac{1}{2}\overline{\psi_c^\ell} \left[ 
		\gamma^0\partial_0  +  i\vec\gamma\cdot\vec k 
		+ a\left( - m_{3/2}^0 + \frac{m_z^2}{\sqrt{3}m^0_{3/2} \Mpl}\delta z \right)
	\right]\psi_c^\ell.
	\label{zLL0}
\end{align}
Here we keep relevant terms in the limit $\delta z \to 0$,
in which we have $p_{\rm SB} / \rho_{\rm SB}\simeq -1,$ $|p_{W}| \ll \rho_{\rm SB}$, etc.
One can see that $\psi^\ell$ obtains a mass of $m_{3/2}^0$ as expected.
The last term is responsible for the Polonyi decay into the longitudinal gravitino pair
and it is clear that there is an enhancement of $\sim (3m_z/m_{3/2}^0)^2$ for the Polonyi decay rate into the longitudinal mode compared with that into the transverse mode.
We have omitted the Hubble mass term since it is suppressed by the ratio $p_{W}/\rho_{\rm SB}$.
This is because, in the limit $\delta z\to 0$,
the $F$ term of $z$ does not contribute to the Hubble expansion 
owing to the requirement that the cosmological constant in the present vacuum is (almost) zero.

\subsection{The case of several chiral superfields}  \label{sec:multigrav}

\subsubsection{General argument} \label{sec:several_general}

Let us consider the multi-field case $\phi_i$ $(i=1,\dots N)$.
We want to express all $N$ chiral fermions $\chi_i$ in terms of the goldstino $v$ and those orthogonal to it.
In this setup, the goldstino is given by \footnote{
	Hereafter, the goldstino $v$ is canonically normalized.  
	Namely, the $v$ in Sec.~\ref{sec:master} is multiplied by $a^{3/2}/\sqrt{\rho_{\text{SB}}}$.
}

\begin{align}
	v_L =\sum_i  \sqrt{\frac{\rho_{\rm SB}^{i}}{\rho_{\rm SB}} }\,v_{iL} \equiv \sum_i \alpha_i v_{iL},
\end{align}
where
\begin{align}
	&v_{iL} \equiv \frac{1}{\sqrt{\rho_{\rm SB}^i}}\left[-F_i^* \chi_{iL} + \dot\phi_i\gamma^0\chi_{iR}\right]
	\equiv \cos\theta_i\,\chi_{iL} + \sin\theta_i\,\gamma^0 \chi_{iR},
\end{align}
and
\begin{align}
	&\rho_{\rm SB}^i \equiv |\dot \phi_i|^2 + |F_i|^2.
\end{align}
Note that $\sum_i \alpha_i^2=1$.
These are left-handed spinors as indicated by the subscript $L$. The right-handed counterparts are similarly given by
\begin{align}
	v_{iR} \equiv \frac{1}{\sqrt{\rho_{\rm SB}^i}}\left[-F_i \chi_{iR} + \dot\phi_i^*\gamma^0\chi_{iL}\right]
	= \cos\theta_i\,\chi_{iR} + \sin\theta_i\, \gamma^0 \chi_{iL},
\end{align}
where we have assumed scalar fields are real.
It is convenient to define the Majorana spinor $\chi_i = \chi_{iL} + \chi_{iR}$, $v=v_L + v_R$, etc.
Then we have
\begin{align}
	v=  \sum_i \alpha_i ( \cos\theta_i + \sin\theta_i\,\gamma^0) \chi_i = \sum_i \alpha_i e^{\gamma^0\theta_i}\,\chi_i.
	\label{vdef}
\end{align}
Similarly there are $N-1$ fermions $v_\perp^I$ $(I=1,\dots N-1)$ orthogonal to the goldstino.
Defining an $N$-component vector $v_i \equiv (v,v_\perp^I)$ $(i=1,\dots,N)$, it can be expressed as
\begin{align}
	v_i
	=\left(O^{T}\right)_{ij} e^{\gamma^0\theta_j}
	\chi_j
	~~~~~\leftrightarrow~~~~~
	\chi_i = e^{-\gamma^0\theta_i}O_{ij} v_j,
\end{align}
where $O$ is an $N\times N$ orthogonal matrix and $O^{T}$ is its transpose, whose first row is determined by Eq.~(\ref{vdef}) as
\begin{align}
	O^{T} = \begin{pmatrix}
	\alpha_1,\dots ,\alpha_N \\ \widetilde O^{T}
	\end{pmatrix},
\end{align}
with $\widetilde O^{T}$ being an $(N-1)\times N$ matrix that satisfies 
$\widetilde O^{T} \widetilde O = {\bf 1}_{(N-1)\times (N-1)}$
from the orthogonality and normalization conditions of $v_\perp^I$.
Also, this matrix fulfills $0 = \alpha_i \widetilde O_{i J}$.

Using these elements defined above, we can express all terms in the Lagrangian including fermions with $v_\perp^I$ in the unitary gauge $v=0$.
The fermion mass term becomes
\begin{align}
	\mathcal L_{f,{\rm mass}} = -\frac{a}{2}\overline{\chi_i}m_{ij} \chi_j 
	= -\frac{a}{2}\overline{v_\perp^I}\left(\widetilde O^{T}_{Ii} e^{\gamma^0\theta_i} m_{ij}e^{-\gamma^0\theta_j}  \widetilde O_{jJ} \right) v_\perp^J.
\end{align}
The fermion kinetic term becomes
\begin{align}
	\mathcal L_{f,{\rm kin}} = -\frac{1}{2}\overline{v_\perp^I} \left[\gamma^0\partial_0 \delta_{IJ} 
	+ \widetilde O^{T}_{I i}e^{\gamma^0\theta_i}  i\left(\vec\gamma\cdot\vec k\right)e^{-\gamma^0\theta_i} \widetilde O_{iJ}
	+ \widetilde O^T_{I i} \left( \der_0 \theta_i \right) \widetilde O_{i J} 
	+ \widetilde O^{T}_{I i}\left(\gamma^0 \partial_0\widetilde O_{iJ}\right) \right] v_\perp^J.
\end{align}
The gradient term seems curious.
However, by taking into account the mixing of $\psiLc$ and $v_{\perp}^I$, it can be diagonalized.
In particular, the off-diagonal gradient term of $\psiLc$ and $v_{\perp}$ comes from $\mathcal L_{\rm mix}$.
The longitudinal gravitino-fermion mixing term (\ref{Lmix}) is rewritten in a simple form as
\begin{align}
	\mathcal L_{\rm mix} = 2\overline{\psiLc}\, i\left(\vec \gamma\cdot\vec k \right)\gamma^0
	\left[ \sum_i \alpha_i \sin\theta_i\,\chi_i \right]
	=2\overline{\psiLc}\, i\left(\vec \gamma\cdot\vec k \right)\gamma^0\alpha_i \sin\theta_i\,e^{-\gamma^0 \theta_i} \widetilde O_{iI} v_\perp^I.
\end{align}
The quantity $\widehat A$ (\ref{Ahat}) is also expressed as
\begin{align}
	\widehat A = -\sum_i\left[ \cos(2\theta_i)+\gamma^0 \sin(2\theta_i) \right] \alpha_i^2 = -\sum_i e^{2\gamma^0\theta_i}\alpha_i^2.
\end{align}
It is easily checked that $|\widehat A|^2=1$ in the single-field case.
Combined with the gravitino kinetic term (\ref{grav_kin}) and $\mathcal L_{\rm mix}$, we obtain a matrix form of the gradient term:
\begin{align}
	\mathcal L_{\rm grad} = -\frac{1}{2}\begin{pmatrix}
		\overline{\psiLc} & \overline {v_{\perp}}
	\end{pmatrix}
	\left( i \vec\gamma\cdot\vec k\right) \widehat{\mathcal{A}}
	\begin{pmatrix}
		\psiLc \\ v_{\perp}
	\end{pmatrix},
\end{align}
where
\begin{align}
	\widehat{\mathcal {A}} =\begin{pmatrix}
		-\widehat {A}^\dagger 
		& \alpha_i\,e^{- 2\gamma^0 \theta_i} \widetilde O_{iJ} \\
		\widetilde O^{T}_{Ii} \alpha_i \,e^{- 2\gamma^0 \theta_i}
		& \widetilde O^{T}_{Ii}e^{-2\gamma^0\theta_i}\widetilde O_{iJ} 
	\end{pmatrix}.
\end{align}
Here and hereafter we suppress the indices of $v_\perp$ for brevity.
This can be written as
\begin{align}
	\widehat{\mathcal A} = O^{T} \,{\rm diag}\left(e^{-2\gamma^0\theta_i}\right) O.
\end{align}
It is easily seen that $|\widehat{\mathcal A}|^2 = 1$ and hence $\widehat{\mathcal A}$ can be regarded as a generalization of $\widehat A$ in the single-field case.
This leads us to define an $N\times N$ matrix $\widehat\theta$ such that $\widehat{\mathcal A} \equiv e^{-2\gamma^0 \widehat\theta}$. Then we have
\begin{align}
	e^{\pm\gamma^0\widehat\theta} = O^{T} \,{\rm diag}\left(e^{\pm\gamma^0\theta_i}\right) O.
\end{align}
Therefore we can diagonalize the gradient term by using the rescaled field
\begin{align}
	\begin{pmatrix}
		{\psiLc}' \\ v_{\perp}'
	\end{pmatrix}
	\equiv e^{-\gamma^0\widehat\theta} \begin{pmatrix}
		{\psiLc} \\ v_{\perp}
	\end{pmatrix}.
	\label{psicellprime}
\end{align}
Using this basis, the kinetic term of ${\psiLc}'$ and $v_\perp'$ is completely diagonal.
On the other hand, this transformation yields the mass mixing between them.
After all, we finally obtain the Lagrangian of the gravitino-fermion system written only by the physical degrees of freedom as
\begin{align}
	\mathcal L= -\frac{1}{2}\begin{pmatrix}
		\overline{{\psiLc}'} & \overline {v_{\perp}'}
	\end{pmatrix}
	\left(\gamma^0\partial_0 + i \vec\gamma\cdot\vec k+ a\mathcal M \right)
	\begin{pmatrix}
		{\psiLc}' \\ v_{\perp}'
	\end{pmatrix},
\end{align}
where
\begin{align}
	\mathcal M = e^{-\gamma^0\widehat\theta}\left[ \gamma^0\frac{\partial}{\partial t} +
	\begin{pmatrix}
		\widehat{m}_{3/2} & 0 \\ 
		0 & \widetilde O^{T}\left(\widehat{m}_f + \dot \theta_i +\gamma^0\dot{\widetilde O}\widetilde O^{T}  \right)\widetilde O
	\end{pmatrix}
	 \right] e^{\gamma^0\widehat\theta},
	 \label{mass_multi}
\end{align}
with
\begin{align}
	\widehat{m}_{f} \equiv e^{\gamma^0\theta_i} m_{ij} e^{-\gamma^0\theta_j}.
\end{align}

For later convenience, we can also estimate $\dot\theta_i$ as
\begin{align}
	\dot\theta_i =\frac{1}{F_i}\left(-\ddot\phi_i+\frac{\dot\phi_i \dot\rho^i_{\rm SB}}{2\rho_{\rm SB}^i} \right)
	= \frac{\partial_{\phi_i} V}{F_i} - \frac{3m_{3/2}\dot\phi_i^2}{\rho_{\rm SB}^i} + \frac{3H \dot\phi_i F_i}{\rho_{\rm SB}^i},
\end{align}
which is of the order of $\sim \mathcal O(m_{\phi_i}) + \mathcal O(m_{3/2}) + \mathcal O(H)$.
It is also conveniently expressed as
\begin{align}
	\dot\theta_i =  \frac{\partial_{\phi_i} V}{F_i} - m_{3/2} - \widehat{m}_{3/2}^i, \label{theta_dot_i}
\end{align}
where we have decomposed
\begin{align}
	\widehat{m}_{3/2} = \sum_i \alpha_i^2 \widehat{m}_{3/2}^i,~~~~~~\widehat{m}_{3/2}^i \equiv \frac{3H p_W^i + m_{3/2}(\rho_{\rm SB}^i + 3p_{\rm SB}^i)}{ 2 \rho_{\rm SB}^i}, \label{mhati_def}
\end{align}
where
\begin{align}
	p_{\rm SB}^i \equiv |\dot\phi_i|^2 - |F_i|^2,~~~p_{W}^i \equiv -(\dot\phi_i^* F_i + \dot \phi_i F_i^*).
\end{align}
Note that $(\rho_{\rm SB}^i)^2 = (p_{\rm SB}^i)^2 + |p_W^i|^2$.
This is a generalization of the single-field case (\ref{theta_single2}).

Note that there are off-diagonal antisymmetric mass terms proportional to $\gamma^0$.
This part can be removed with a transformation by an orthogonal matrix as described in Sec.\,3.2 of Ref.~\cite{Nilles:2001fg}, and hence does not contribute to the mass eigenvalues.

\subsubsection{Two-field example}

Now let us consider the two-field case, which is used in Sec.~\ref{sec:single}.
In this case, after removing the goldstino $v$ in the unitary gauge, there remains only one fermion $v_\perp$ other than the gravitino.
The orthogonal matrix $O$  and $\widetilde O$ are easily found to be
\begin{align}
	O =\begin{pmatrix}
		\alpha_1 & -\alpha_2 \\
		\alpha_2 & \alpha_1
	\end{pmatrix},~~~~~~~
	\widetilde O = \begin{pmatrix}
		-\alpha_2 \\
		\alpha_1
	\end{pmatrix}.
\end{align}
The goldstino and the fermionic degree of freedom orthogonal to the goldstino $v$ are,
\begin{align}
	\begin{pmatrix}
	v \\ v_\perp
	\end{pmatrix}
	= 
	\begin{pmatrix}
	\alpha_1 & \alpha_2 \\ -\alpha_2 & \alpha_1
	\end{pmatrix}
	\begin{pmatrix}
	e^{\gamma^0\theta_1} & 0 \\ 0 & e^{\gamma^0\theta_2}
	\end{pmatrix}
	\begin{pmatrix}
	\chi_1 \\ \chi_2
	\end{pmatrix},
\end{align}
which is inversely solved as
\begin{align}
	\begin{pmatrix}
	\chi_1 \\ \chi_2
	\end{pmatrix}
	=
	\begin{pmatrix}
	\alpha_1e^{-\gamma^0\theta_1} & -\alpha_2e^{-\gamma^0\theta_1} \\ \alpha_2e^{-\gamma^0\theta_2} & \alpha_1e^{-\gamma^0\theta_2}
	\end{pmatrix}
	\begin{pmatrix}
	v \\ v_\perp
	\end{pmatrix}
	=
	\begin{pmatrix}
	-\alpha_2e^{-\gamma^0\theta_1}\,v_\perp \\   \alpha_1e^{-\gamma^0\theta_2}\,v_\perp
	\end{pmatrix}.
\end{align}
In the last equality, we have taken the unitary gauge $v=0$.
First, the fermion mass term becomes
\begin{align}
	&\mathcal L_{f,{\rm mass}} =-\frac{1}{2}\bar\chi_i a m_{ij} \chi_j 
	= -\frac{1}{2}\overline{v_\perp} a m_f v_\perp, \\
	&m_f\equiv \alpha_2^2m_{11} + \alpha_1^2m_{22} - 2\alpha_1\alpha_2 m_{12}\cos(\theta_1-\theta_2).  \label{mf}
\end{align}
We can substitute this expression into the fermion kinetic term:
\begin{align}
	\mathcal L_{f,{\rm kin}} &= -\frac{1}{2}\left( \bar\chi_1\slashed{\partial}\chi_1 + \bar\chi_2\slashed{\partial}\chi_2  \right) \nonumber\\
	&= -\frac{1}{2} \left[ \overline{v_\perp}\gamma^0\partial_0 v_\perp  + \overline{v_\perp} 
	\left( \alpha_2^2 e^{2\gamma^0\theta_1} + \alpha_1^2 e^{2\gamma^0\theta_2}  \right) \gamma^i\partial_i v_\perp \right]
	-\frac{1}{2}\overline{v_\perp} a\left(\alpha_2^2\dot\theta_1 + \alpha_1^2\dot\theta_2\right) v_\perp.
\end{align}
The matrix of the gradient term is given by
\begin{align}
	-\frac{1}{2}  
	\begin{pmatrix}
	\overline{\psiLc} & \overline{v_\perp}
	\end{pmatrix}
	i \left(\vec\gamma\cdot\vec k \right) \widehat {\mathcal A}
	\begin{pmatrix}
	\psiLc \\ v_\perp
	\end{pmatrix},
\end{align}
where $\widehat {\mathcal A}$ is a $2\times 2$ matrix defined by
\begin{align}
	\widehat {\mathcal A} =
	\begin{pmatrix}
	 \alpha_1^2 e^{-2\gamma^0\theta_1} + \alpha_2^2 e^{-2\gamma^0\theta_2} &
	 -\alpha_1\alpha_2\left( e^{-2\gamma^0\theta_1} - e^{-2\gamma^0\theta_2} \right) \\
	 -\alpha_1\alpha_2\left( e^{-2\gamma^0\theta_1} - e^{-2\gamma^0\theta_2} \right) &
	 \alpha_2^2 e^{-2\gamma^0\theta_1} + \alpha_1^2 e^{-2\gamma^0\theta_2}
	\end{pmatrix}.
\end{align}
As noted earlier, the matrix $\widehat {\mathcal A}$ may be regarded as a generalization of $\widehat A$.
We can express $\widehat {\mathcal A}$ as $\widehat {\mathcal A} = e^{-2\gamma^0 \widehat \theta}$ with $\widehat\theta$ being a real symmetric matrix.
Therefore we can diagonalize the gradient term by redefining the fields as in Eq.~(\ref{psicellprime}).
Explicitly,
\begin{align}
\widehat{\theta} = \begin{pmatrix}
\alpha_1^2 \theta_1 + \alpha_2^2 \theta_2 & -\alpha_1 \alpha_2 (\theta_1 - \theta_2) \\
- \alpha_1 \alpha_2 (\theta_1 - \theta_2) & \alpha_2^2 \theta_1 + \alpha_1^2 \theta_2 
 \end{pmatrix},
 \end{align}
and
\begin{align}
	e^{\gamma^0\widehat\theta} =
	\begin{pmatrix}
	 \alpha_1^2 e^{\gamma^0\theta_1} + \alpha_2^2 e^{\gamma^0\theta_2} &
	 -\alpha_1\alpha_2\left( e^{\gamma^0\theta_1} - e^{\gamma^0\theta_2} \right) \\
	 -\alpha_1\alpha_2\left( e^{\gamma^0\theta_1} - e^{\gamma^0\theta_2} \right) &
	 \alpha_2^2 e^{\gamma^0\theta_1} + \alpha_1^2 e^{\gamma^0\theta_2}
	\end{pmatrix}.
\end{align}
After the diagonalization of the gradient term, the full fermionic Lagrangian becomes
\begin{align}
	\mathcal L_{f} = -\frac{1}{2}\begin{pmatrix}
	\overline{{\psiLc}'} & \overline{{v_\perp}'} 
	\end{pmatrix}
	\left[
	\gamma^0\partial_0 + i\vec\gamma\cdot\vec k
	+a\mathcal M
	\right]
	\begin{pmatrix}
	{\psiLc}' \\ {v_\perp}'
	\end{pmatrix},
\end{align}
where $\mathcal M$ denotes the mass matrix given by
\begin{align}
	&\mathcal M = 
	e^{-\gamma^0\widehat\theta}
	\left[
	\gamma^0\frac{\partial}{\partial t} + 
	\begin{pmatrix}
	\widehat{m}_{3/2} & 0\\  
	0 & \alpha_2^2\dot\theta_1 + \alpha_1^2\dot\theta_2 + m_f
	\end{pmatrix} 
	\right]
	e^{\gamma^0\widehat\theta}.
	\label{M_2field}
\end{align}
Clearly, in the single-field dominance limit $\alpha_1 \gg \alpha_2$ or $\alpha_2 \gg \alpha_1$, 
the off-diagonal elements are suppressed by the ratio $\alpha_2/\alpha_1$ or $\alpha_1/\alpha_2$.
Therefore the mixing between $\psiLc$ and $v_\perp$ is also suppressed by this ratio,
and in this limit we effectively recover the single field case studied in Sec.~\ref{sec:singlegrav}.

An explicit expression for the mass matrix $\mathcal M=\mathcal M_1+\mathcal M_2$ is given by
\begin{align}
\mathcal{M}_1 \equiv& e^{-\gamma^0 \widehat{\theta}} \gamma^0 \frac{\partial}{\partial t} e^{\gamma^0 \widehat{\theta}}, \\
(\mathcal{M}_1)_{11} =& - \alpha_1^2 \dot\theta_1 - \alpha_2^2  \dot\theta_2 -2 \alpha_1 \dot\alpha_1 \sin(\theta_1 - \theta_2), \nonumber\\
(\mathcal{M}_1)_{12}=&   \alpha_1 \alpha_2  (\dot\theta_1 -\dot\theta_2 ) +(\alpha_1 \dot\alpha_2 + \alpha_2 \dot\alpha_1) \sin(\theta_1 - \theta_2) - (\alpha_1 \dot\alpha_2 - \alpha_2 \dot\alpha_1)(1-\cos (\theta_1-\theta_2))\gamma^0
, \nonumber\\
(\mathcal{M}_1)_{21}=&  \alpha_1 \alpha_2  (\dot\theta_1 - \dot\theta_2 )+ (\alpha_1 \dot\alpha_2 + \alpha_2 \dot\alpha_1) \sin(\theta_1 - \theta_2) + (\alpha_1 \dot\alpha_2 - \alpha_2 \dot\alpha_1)(1-\cos (\theta_1-\theta_2))\gamma^0
, \nonumber\\
(\mathcal{M}_1)_{22}=&  - \alpha_2^2 \dot\theta_1 - \alpha_1^2 \dot\theta_2- 2 \alpha_2 \dot\alpha_2 \sin(\theta_1 - \theta_2),\nonumber
\end{align}
and
\begin{align}
\mathcal{M}_2 \equiv & e^{-\gamma^0 \widehat{\theta}} \text{diag} \left(\widehat{m}_{3/2}, ~m_f + \alpha_2^2 \dot\theta_1 + \alpha_1^2 \dot\theta_2  \right)  e^{\gamma^0 \widehat{\theta}}, \\
(\mathcal{M}_2)_{11} =& \left(1 - 2\alpha_1^2 \alpha_2^2 \left(1-\cos (\theta_1 - \theta_2) \right) \right)\widehat{m}_{3/2} 
+2 \alpha_1^2 \alpha_2^2 \left(1-\cos (\theta_1 - \theta_2) \right)(  m_f  + \alpha_2^2 \dot\theta_1 + \alpha_1^2 \dot\theta_2 ),\nonumber \\
(\mathcal{M}_2)_{12} =& \alpha_1 \alpha_2 \left((\alpha_1^2 -\alpha_2^2) \left( 1 - \cos \left( \theta_1-\theta_2 \right) \right)  + \gamma^0 \sin \left(\theta_1 - \theta_2 \right) 
\right) 
  \left(  ( m_f  + \alpha_2^2\dot \theta_1 + \alpha_1^2 \dot\theta_2) -\widehat{m}_{3/2}  \right), \nonumber\\
(\mathcal{M}_2)_{21} =& \alpha_1 \alpha_2 \left((\alpha_1^2 -\alpha_2^2) \left( 1 - \cos \left( \theta_1-\theta_2 \right) \right)  - \gamma^0 \sin \left(\theta_1 - \theta_2 \right) 
\right) 
 \left(  ( m_f  + \alpha_2^2 \dot\theta_1 + \alpha_1^2 \dot\theta_2) -\widehat{m}_{3/2} \right), \nonumber\\
(\mathcal{M}_2)_{22} =& 2\alpha_1^2 \alpha_2^2 \left( 1- \cos (\theta_1 - \theta_2) \right)\widehat{m}_{3/2} 
+ \left( 1- 2 \alpha_1^2 \alpha_2^2 \left( 1- \cos (\theta_1 - \theta_2) \right) \right) (m_f  + \alpha_2^2 \dot\theta_1 + \alpha_1^2 \dot\theta_2). \nonumber
\end{align}
As mentioned above, the terms proportional to $\gamma^0$ do not affect the mass eigenvalues.

\section{Gravitino production in single-superfield inflation}  \label{sec:single}

Let us consider the two-field case, in which there are two chiral superfields: inflaton $\phi$ and Polonyi $z$.
Their fermionic components are denoted by $\widetilde\phi$ and $\widetilde z$, respectively.
The K\"ahler and superpotential are assumed to be
\begin{align}
	&K = -\frac{1}{2}(\phi - \phi^\dagger)^2 + |z|^2 - \frac{|z|^4}{\Lambda^2}, \\
	&W= \frac{1}{2}m_\phi \phi^2 + \mu^2 z + W_0.  
\end{align}
Although we have imposed an approximate shift symmetry $\phi\to\phi+c$ with $c$ being a real constant, almost all the following arguments 
do not depend on this specific choice of the K\"ahler potential as long as $K_{\phi\bar\phi} \simeq 1$.
Without loss of generality, we can take all the coupling  constants real and positive.
In the present vacuum, $\phi=0$ and $\left< z\right> \simeq 2\sqrt{3}\Mpl (m^{0}_{3/2}/m_z)^2$
where $m_z^2 = 12(m^{0}_{3/2}\Mpl/\Lambda)^2$.
Here $m^{0}_{3/2} \simeq \mu^2/\sqrt{3}\Mpl \simeq W_0/\Mpl^2$ is the gravitino mass in the present universe.
We focus on the case with the $\mathbb{Z}_2$ symmetry in which there is no linear term in the K\"ahler potential and the inflaton oscillates around $\phi=0$.
The case without the $\mathbb{Z}_2$ symmetry will be discussed in Sec.~\ref{sec:comment}.

One should note that this theory does not lead to successful chaotic inflation:
the simple power-law behavior of the inflaton potential $V \simeq m_\phi^2 |\phi|^{2}$ is ensured only at the sub-Planckian
field range $|\phi| \lesssim \Mpl$.
Therefore we need to carefully choose the K\"ahler and/or superpotential to modify the potential at large field value for 
successful inflation~\cite{Goncharov:1983mw,Ketov:2014qha,Ketov:2014hya,Linde:2014hfa, Roest:2015qya, Linde:2015uga, Scalisi:2015qga, Ketov:2016gej, Ferrara:2016vzg}.
However, we are interested in the behavior after inflation with sub-Planckian field value,
and hence such modifications on the potential at large field value do not significantly affect the following discussion.

\begin{figure}[t]
\begin{center}
\includegraphics[scale=0.8]{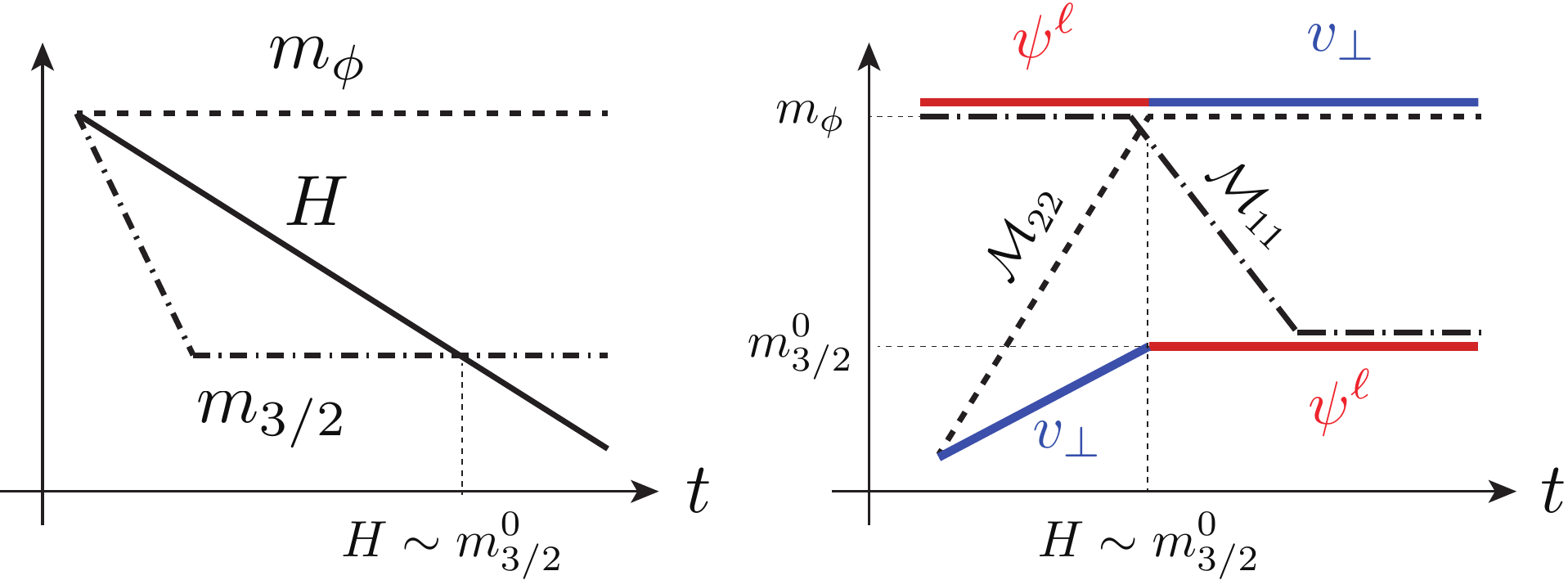}
\end{center}
\caption {\small
\textbf{Left}: Time dependence of $m_\phi, H$ and $m_{3/2}$ in single superfield inflation model.
\textbf{Right}: Time evolution of mass eigenvalues of $(\psiL, v_\perp)$ are shown by thick solid lines.
The \textcolor{dark_red}{red} (\textcolor{dark_blue}{blue}) 
segments show that the main composition of the mass eigenstate is $\psi^\ell$ ($v_\perp$).
Dashed and dot-dashed lines show $\mathcal M_{22}$ and $\mathcal M_{11}$, respectively.
}
\label{fig:1}
\end{figure}

\subsection{Dynamics} \label{sec:dyn_single}

First let us briefly summarize the scalar dynamics in the present model.
The inflaton dynamics is so simple that the inflaton $\phi$ just oscillates around the origin $\phi=0$
along the axis of ${\rm Re}\,\phi$ after inflation.
The oscillation amplitude $\phi_{\rm amp}$ decreases as $\phi_{\rm amp} \propto a^{-3/2} \sim t^{-1}$
until it completely decays and the universe is reheated.

The $z$ dynamics is slightly complicated.
The oscillation of $z$ is directly induced by the inflaton dynamics through the potential term
\begin{align}
	V \supset \frac{\mu^2m_\phi \phi}{\Mpl^2}\left(K_{\bar\phi}-\phi\right)z^* + {\rm h.c.} \simeq 
	-3 m_{3/2}^0 H \frac{|\phi|^2 z^*}{\phi_{\rm amp}}+ {\rm h.c.}
	\label{V_mix_eff}
\end{align}
Although this term is not regarded as a ``mixing'' between $\phi$ and $z$ formally,
it inevitably induces a coherent oscillation in the $z$ direction when $\phi$ has a finite oscillation amplitude.
In that sense, this may be viewed as an effective mixing term between $\phi$ and $z$, with the mixing angle of\footnote{
	As is clear from Eq.~\eqref{V_mix_eff}, the mixing term of Eq.~\eqref{eq:induced_mixing} vanishes for a quadratic superpotential $W \propto \phi^2$ and the minimal K\"ahler potential $K=|\phi|^2$.
}
\begin{align}
	\theta_{\phi z} \sim \begin{cases}
		\displaystyle\frac{m_{3/2}^0 H}{m_\phi^2} & {\rm for}~~m_z< m_\phi, \\ ~&~\\
		\displaystyle\frac{m_{3/2}^0 H}{m_z^2}  & {\rm for}~~m_z > m_\phi.
	\end{cases}
	\label{eq:induced_mixing}
\end{align}
As shown in App.~\ref{sec:dyn}, the induced amplitude of $z$ can be estimated as
$z_{\rm amp}^{\rm (ind)}\sim \theta_{\phi z} \phi_{\rm amp}$.
This nonzero amplitude of $z$ may contribute to the longitudinal gravitino production as will be shown later.
Note that this ``induced'' oscillation of $z$ always exists even for $m_z \gg H_{\rm inf}$.
It can be interpreted as a result of tilted axis of the oscillation on $(\phi,z)$ plane due to the effective mixing term (\ref{V_mix_eff}),
hence it is just a small mixture of the $z$ into dominantly $\phi$ oscillation.

On the other hand, if $m_z \ll H_{\rm inf}$, there is another oscillation mode which dominantly consists of 
the $z$ direction, which occurs at $H \sim m_z$. 
This is roughly the oscillation along the light mass eigenstate, which mostly consists of $z$.
If $m_z\gg H_{\rm inf}$, the adiabatic suppression of the coherent oscillation works~\cite{Linde:1996cx,Nakayama:2011wqa}
and the non-induced oscillation is safely neglected.
Therefore, taking account of the Hubble expansion, the oscillation amplitude is given by~\cite{Nakayama:2012hy}
\begin{align}
	z_{\rm amp}^{\rm (non-ind)} \simeq \begin{cases}
		\displaystyle\left< z\right>\left(\frac{a(t_{\rm osc})}{a(t)}\right)^{3/2} = \frac{2\sqrt{3} (m_{3/2}^0)^2 \Mpl}{m_z^2}\left(\frac{a(t_{\rm osc})}{a(t)}\right)^{3/2}& {\rm for}~~~m_z \ll H_{\rm inf},\\~&~\\
		\displaystyle 0  & {\rm for}~~~m_z \gg H_{\rm inf},
	\end{cases}
	\label{z_amp_light}
\end{align}
where $a(t_{\rm osc})$ denotes the scale factor at $H=m_z$.
The resultant oscillation amplitude is the sum of the induced one and the non-induced one:
\begin{align}
	z_{\rm amp} \sim z_{\rm amp}^{\rm (ind)} + z_{\rm amp}^{\rm (non-ind)}.
\end{align}
The effect of the non-induced one (\ref{z_amp_light}) on the gravitino production was considered in Refs.~\cite{Nakayama:2012hy,Evans:2013nka}
and we mainly focus on the effect of the induced one.

\subsection{Transverse component}  \label{sec:trans_single}

First let us evaluate the transverse gravitino production rate.
The Lagrangian is given by Eq.~(\ref{Ltrans}), and the rapid oscillation of the mass $m_{3/2}$ contributes to the gravitino production.
The oscillating part of the gravitino mass is given by $m_{3/2} \simeq m_\phi\phi^2/2\Mpl^2$.
A schematic picture of the time evolution of $m_{3/2},H$ and $m_\phi$ is shown in the left panel of Fig.~\ref{fig:1}.
Hence, as shown in App.~\ref{sec:fermion}, we obtain the effective ``annihilation'' rate of the inflaton as
\begin{align}
	\Gamma(\phi\phi\to\psiT \psiT) \simeq \frac{\mathcal C}{16\pi} \frac{\phi_{\rm amp}^2}{\Mpl^2} \frac{m_\phi^3}{\Mpl^2}
	\simeq \frac{3\mathcal C}{16\pi}\frac{H^2 m_\phi}{\Mpl^2},
\end{align}
where $\mathcal C$ is a numerical constant of order unity.
We take $\mathcal C = 1$ as a reference [See Eq.~\eqref{gamma_general}].
Since $m_\phi \gtrsim m_{3/2}$ always holds after inflation, this process is kinematically accessible.\footnote{
	Gravitinos produced in this way become more and more relativistic as the gravitino mass $m_{3/2}$ becomes smaller 
	due to the cosmic expansion. For most cases, it eventually becomes non-relativistic before it decays.
}
This production rate is comparable to the gravitational particle production rate of a minimal scalar field~\cite{Ford:1986sy,Ema:2015dka},
although the gravitino is actually conformal in the massless limit.
Note that since the transverse mode does not have mixing with any other fermion,
the transverse mode produced in any epoch eventually becomes the present gravitino.

Since the amplitude $\phi_{\rm amp}$ is rapidly decreasing with respect to the cosmic expansion, 
the transverse gravitino production becomes less and less effective as time goes on.
This is shown by checking that the gravitino number density produced per Hubble time,
\begin{align}
	\frac{\rho_\phi}{m_\phi} \frac{\Gamma(\phi\phi\to\psiT\psiT) }{H},
\end{align}
decreases faster than $a^{-3}$, or $\Gamma(\phi\phi\to\psiT\psiT)$ decreases faster than $H$.
Thus the dominant contribution comes from those created within a few inflaton oscillation just after inflation.
The present gravitino abundance is estimated as 
\begin{align}
	\frac{n_{3/2}^{(t)}}{s} &\simeq \left(\frac{\Gamma(\phi\phi\to \psiT\psiT)}{H}\right)_{H=H_{\rm inf}} \frac{3T_{\rm R}}{4m_\phi}
	\simeq
	\frac{9\mathcal C}{64\pi}\frac{H_{\rm inf} T_{\rm R}}{\Mpl^2} \nonumber\\
	&\simeq 8\times10^{-16} 
	\mathcal C
	\left( \frac{H_{\rm inf}}{10^{13}\,{\rm GeV}} \right)
	\left( \frac{T_{\rm R}}{10^{10}\,{\rm GeV}} \right),
	\label{trans_single}
\end{align}
where $T_{\rm R}$ is the reheating temperature and $H_{\rm inf}$ denotes the Hubble scale at the end of inflation.
This is about three orders of magnitude smaller than the contribution from thermal production.

\subsection{Longitudinal component}  \label{sec:long_single}

Next let us consider the production of longitudinal gravitino.
As already mentioned in Sec.~\ref{sec:grav}, we must be careful on the mixing with fermions
to correctly treat the production and evolution of the longitudinal gravitino.

In the present case, there is one physical fermionic degree of freedom $v_\perp$ in the matter sector,
and it mixes with the longitudinal gravitino.
The goldstino is given by
\begin{align}
	v = \frac{1}{\sqrt{\rho_{\rm SB}}}\left[ \sqrt{\rho_{\rm SB}^\phi}v_\phi 
	+ \sqrt{\rho_{\rm SB}^z} v_z  \right],
\end{align}
where
\begin{align}
	&v_\phi \equiv \frac{1}{\sqrt{\rho_{\rm SB}^\phi}}\left(-F_\phi + \dot\phi\gamma^0\right)\widetilde\phi,
	~~~\rho_{\rm SB}^\phi \equiv |\dot \phi|^2 + |F_\phi|^2,\\
	&v_z \equiv  \frac{1}{\sqrt{\rho_{\rm SB}^z}}\left(-F_z + \dot z\gamma^0\right)\widetilde z,
	~~~\rho_{\rm SB}^z \equiv |\dot z|^2 + |F_z|^2.
\end{align}
Note again that we have assumed that scalar fields have real values: this is justified since we have taken all model parameters 
and also the initial condition during inflation real.
In the unitary gauge, $v$ is set to be zero.
The remaining fermionic degree of freedom is that orthogonal to the goldstino,
\begin{align}
	v_{{\perp}} = \frac{1}{\sqrt{\rho_{\rm SB}}}\left[- \sqrt{\rho_{\rm SB}^z}v_\phi 
	+ \sqrt{\rho_{\rm SB}^\phi} v_z \right].
\end{align}
As a rough estimation, we have 
\begin{align}
	&|\dot\phi| \sim |F_\phi| \sim m_\phi \phi_{\rm amp}\sim H\Mpl,~~~|F_z| \sim \mu^2 \sim m_{3/2}^0 \Mpl \gg |\dot z|,\\
	&m_{3/2}\simeq \frac{m_\phi\phi^2}{2\Mpl^2} + m_{3/2}^0 \sim H\frac{\phi_{\rm amp}}{\Mpl} + m_{3/2}^0,
\end{align}
where $\phi_{\rm amp}$ denotes the amplitude of inflaton oscillation.
Thus we have
\begin{align}
	v \sim \begin{cases}
		\widetilde \phi     &  {\rm for}  ~~H \gtrsim m_{3/2}^0\\
		\widetilde z &  {\rm for} ~~H \lesssim m_{3/2}^0
	\end{cases},~~~~~~
	v_{\perp} \sim \begin{cases}
		\widetilde z     &  {\rm for}  ~~H \gtrsim m_{3/2}^0\\
		\widetilde \phi &  {\rm for} ~~H \lesssim m_{3/2}^0
	\end{cases}.
\end{align}
Hence $z$ begins to dominate the SUSY braking at $H \sim m_{3/2}^0$ while the gravitino mass becomes dominated by the 
present value earlier.
Therefore, whenever $H \lesssim m_{3/2}$, we have $m_{3/2} = m_{3/2}^0$.

The Lagrangian of the longitudinal gravitino and fermion system is found in Sec.~\ref{sec:multigrav} as
\begin{align}
	\mathcal L_{f} = -\frac{1}{2}\begin{pmatrix}
	\overline{{\psiLc}'} & \overline{{v_\perp}'} 
	\end{pmatrix}
	\left[
	\gamma^0\partial_0 + i\vec\gamma\cdot\vec k
	+a\mathcal M
	\right]
	\begin{pmatrix}
	{\psiLc}' \\ {v_\perp}'
	\end{pmatrix},
\end{align}
where $\mathcal M$ denotes the mass matrix defined in Eq.~(\ref{M_2field}).
Identifying $1 \to \phi$ and $2\to z$ in the notation of Sec.~\ref{sec:multigrav},
 $\mathcal M$ is roughly estimated as
\begin{align}
	\mathcal M \simeq 
	\begin{pmatrix}
	 m_\phi & (m_{3/2}^0/H) m_\phi \\
	 (m_{3/2}^0/H) m_\phi & m_f
	\end{pmatrix}
	\simeq 
	\begin{pmatrix}
	 m_\phi & (m_{3/2}^0/H) m_\phi \\
	 (m_{3/2}^0/H) m_\phi & (m_{3/2}^0/H)^2 m_\phi
	\end{pmatrix},
\end{align}
for $H\gtrsim m_{3/2}^0$ and
\begin{align}
	\mathcal M \simeq 
	\begin{pmatrix}
	  (H/m_{3/2}^0)^2 m_\phi & (H/m_{3/2}^0) m_\phi \\
	 (H/m_{3/2}^0) m_\phi & m_f
	\end{pmatrix}
	\simeq 
	\begin{pmatrix}
	  (H/m_{3/2}^0)^2 m_\phi & (H/m_{3/2}^0) m_\phi \\
	 (H/m_{3/2}^0) m_\phi & m_\phi
	\end{pmatrix},
\end{align}
for $H\lesssim m_{3/2}^0$, where the first $\simeq$'s involve a similarity transformation of the matrix.
The analysis in App.~\ref{sec:mass} shows that the determinant of $\mathcal M$ is
\begin{align}
	\det\mathcal M \simeq - m_{3/2} m_f \simeq - \alpha_2^2 m_{\phi}m_{3/2},
\end{align}
where $m_{f}$ is defined in Eq.~(\ref{mf}), and $\alpha_2 \sim \min [ m_{3/2}^0 / H, 1]$ whereas the trace of $\mathcal M$ is 
\begin{align}
\tr \mathcal M \simeq m_{\phi}.
\end{align}
Therefore the mass eigenvalues read
\begin{align}
	\begin{pmatrix}
	 m_{\rm heavy}, &
	 m_{\rm light}
	\end{pmatrix}
	\simeq 
	\begin{cases}
	\begin{pmatrix}
	 m_{\phi}, &
	 -m_{3/2}(m_{3/2}^0/H)^2
	\end{pmatrix}
	&{\rm for}~~H\gtrsim m_{3/2}^0
	\\
	\begin{pmatrix}
	 m_{\phi}, &
	 -m_{3/2}
	\end{pmatrix}
	&{\rm for}~~H\lesssim m_{3/2}^0
	\end{cases}.
\end{align}
A more detailed derivation of this result is given in App.~\ref{sec:mass}.
The right panel of Fig.~\ref{fig:1} shows the schematic picture of the time evolution of the mass matrix structure. 
What we regard as the ``present gravitino'' is the light eigenstate at the late epoch $(H \ll m_{3/2}^0)$ which is mainly composed of $\psiLc$.
The heavy mass eigenstate at the late epoch ($v_\perp \sim \widetilde\phi$), on the other hand, is regarded as the inflatino.
In principle, there are two contributions to the final gravitino abundance.
\begin{itemize}
\item The heavy mass eigenstate produced at the early epoch $(H \gg m_{3/2}^0)$ converts into the light mass eigenstate (the present gravitino)
at the late epoch.
\item Almost all the light mass eigenstate produced at the early or late epoch becomes the present gravitino.
\end{itemize}
However, the former contribution is negligible for the following reason.
From the structure of the mass matrix, the mixing angle between $\psiLc$ and $v_\perp$ is about
\begin{align}
	\theta_{\psiLc-v_\perp} \sim \min\left[ \frac{m_{3/2}^0}{H} ,\frac{H}{m_{3/2}^0}\right],
\end{align}
and we always find $\dot\theta_{\psiLc-v_\perp} \sim H \theta_{\psi_c^\ell-v_\perp} \ll m_\phi$,
meaning that the time evolution of the mixing is always adiabatic.
Therefore we can neglect the conversion of the heavy mass eigenstate produced at the early epoch (dominantly $\psiLc$)
into the light mass eigenstate at the late epoch (again dominantly $\psiLc$).\footnote{
	The produced heavy mass eigenstate eventually becomes the inflatino $(v_\perp \simeq \widetilde\phi)$~\cite{Nilles:2001ry,Nilles:2001fg}.
}
Below we analyze the production of the light mass eigenstate and compute the final longitudinal gravitino abundance.
Also we will comment on the production of the heavy mass eigenstate.

\subsubsection*{Early epoch : $H > m_{3/2}^0$}

At the early epoch $H> m_{3/2}^0$, the light mass eigenstate is almost $v_\perp' (\sim \widetilde z)$.
It eventually becomes the present longitudinal gravitino at the later epoch as seen in Fig.~\ref{fig:1}.
Thus we estimate the production rate of $v_\perp$ in this epoch to derive the final longitudinal gravitino abundance.
Taking the limit $\alpha_2\ll 1$ in the general two-field Lagrangian (\ref{M_2field}), we have
\begin{align}
	\mathcal L_{v_\perp} \simeq -\frac{1}{2}\overline{v_\perp'} \left[ 
		\gamma^0\partial_0  +  i\vec\gamma\cdot\vec k + a m_{\rm light}
	\right] v_\perp',
\end{align}
where
\begin{align}
	m_{\rm light} \simeq - m_{3/2} \left(\frac{m_{3/2}^0}{H}\right)^2  + \frac{m_z^2}{\sqrt{3} m_{3/2}^0 \Mpl}\delta z.  \label{mlight_early}
\end{align}
A very important note here is that $m_{3/2}$ appearing in $m_{\rm light}$ does not contain an oscillating part of the order of 
$p_{\rm SB}/\rho_{\rm SB} \sim \mathcal O(1)$ despite that $\widehat{m}_{3/2}$ contains such a violently oscillating part.
Actually a careful calculation of the mass eigenvalues of the full mass matrix in App.~\ref{sec:mass} shows that such a violently oscillating term
cancels out in the light mass eigenvalue $m_{\rm light}$ (but not for $m_{\rm heavy}$).
Therefore, the $\phi$ dependence of $m_{\rm light}$ is at most $m_\phi\phi^2/\Mpl^2$ which may be included in $m_{3/2}$,
leading to the upper bound on the production rate as
\begin{align}
	\Gamma(\phi\phi\to v_\perp v_\perp) 
	\lesssim \frac{\mathcal C}{16\pi}\left( \frac{m_{3/2}^0}{H} \right)^4 \frac{\phi_{\rm amp}^2}{\Mpl^2} \frac{m_\phi^3}{\Mpl^2}
	\simeq \frac{3\mathcal C}{16\pi}\left(\frac{m_{3/2}^0}{H} \right)^2 \frac{(m_{3/2}^0)^2 m_\phi}{\Mpl^2}.
\end{align}
It is suppressed by a factor $(m_{3/2}^0/H)^4$ compared with the transverse gravitino production rate.
The origin of this suppression is the fact that the Polonyino $\widetilde{z}$ is massless in the global SUSY limit.  Although it is absorbed by gravitino to make it massive by the super-Higgs mechanism in the late epoch, goldstino is mainly composed of inflatino in the early epoch and the Polonyino remains light.

Note also that $z$ oscillation also contributes to the production.
As long as we are concerned with the induced oscillation explained in Sec.~\ref{sec:dyn_single},
its contribution to the light mass eigenstate $(\sim v_\perp)$ production rate is at most the same order of the upper bound from $\phi$ annihilation mentioned above.\footnote{
	The production rate of high momentum scalar components of $z$ through the inflaton oscillation is also the same order~\cite{Nakayama:2012hy}.
}

\subsubsection*{Late epoch : $H < m_{3/2}^0$}

The longitudinal gravitino $\psiL$ produced at $H<m_{3/2}^0$ is essentially the same as the present gravitino.
Thus, we can estimate the final longitudinal gravitino abundance by simply evaluating the production rate of $\psiL$ at $H<m_{3/2}^0$.
Taking the limit $\alpha_2\ll \alpha_1$ in the general two-field Lagrangian (\ref{M_2field}), we find the
Lagrangian of $\psiL$ same as that discussed in Sec.~\ref{sec:Polonyi}: 
\begin{align}
	\mathcal L_\ell = -\frac{1}{2}\overline{{\psiLc}'} \left[ 
		\gamma^0\partial_0  +  i\vec\gamma\cdot\vec k + am_{\rm light}
	\right]{\psiLc}',
	\label{zLL}
\end{align}
where 
\begin{align}
	m_{\rm light} \simeq - m_{3/2} + \frac{m_z^2}{\sqrt{3} m_{3/2}^0 \Mpl} \delta z.  \label{mlight}
\end{align}
Again, it is important to note that $m_{3/2}$ in the expression of $m_{\rm light}$ does not contain a violently oscillating part
and the $\phi$ dependence of $m_{\rm light}$ is at most $m_\phi\phi^2/\Mpl^2$.\footnote{ \label{ftn:m32hat_vs_m32}
	Note that $m_{3/2}$ and $\widehat{m}_{3/2}$ are the same order for $H < m_{3/2}^0$,
	but the magnitude of their oscillating part is significantly different.
	The relative oscillating amplitude of $m_{3/2}$ is $\sim H^2/(m_{3/2}^0 m_\phi)$
	while that of $\widehat{m}_{3/2}$ is $\sim (H/m_{3/2}^0)^2$.
	Thus the latter is $m_\phi/m_{3/2}^0$ times larger than the former. 
	Therefore it is very important to figure out what exactly is the ``gravitino mass'' of the light mass eigenvalue.
}
Thus we obtain an upper bound on the production rate of $\psiL$ same as that of the transverse gravitino  as
\begin{align}
	\Gamma(\phi\phi\to\psiL\psiL) 
	\lesssim \frac{\mathcal C}{16\pi} \frac{\phi_{\rm amp}^2}{\Mpl^2} \frac{m_\phi^3}{\Mpl^2}
	\simeq \frac{3\mathcal C}{16\pi}\frac{H^2 m_\phi}{\Mpl^2}.
\end{align}
This is a rapidly time decreasing function and becomes less efficient at the later epoch.
 
 We again note that the contribution of the induced $z$ oscillation to the longitudinal gravitino production is
 at most the same order of the upper bound from $\phi$ annihilation mentioned above.

\subsubsection*{Abundance of longitudinal gravitino}

Combining the results of production rate at the early and late epochs, 
we find that the longitudinal gravitino production is most efficient around $H \sim m_{3/2}^0$.
Thus the abundance of longitudinal gravitino is found to be
\begin{align}
	\frac{n_{3/2}^{(\ell)}}{s} 
	&\lesssim \left(\frac{\Gamma(\phi\phi\to \psiL\psiL)}{H}\right)_{H=m_{3/2}^0} \frac{3T_{\rm R}}{4m_\phi}
	= \frac{9\mathcal C}{64\pi} \frac{m_{3/2}^0 T_{\rm R}}{\Mpl^2} \nonumber\\
	& \simeq 8\times 10^{-23}\, 
	\mathcal C
	\left( \frac{m_{3/2}^0}{10^{6}\,{\rm GeV}} \right)
	\left( \frac{T_{\rm R}}{10^{10}\,{\rm GeV}} \right).
	\label{Y_single}
\end{align}
It is much smaller than the corresponding transverse gravitino abundance.
The dotted line in the left panel of Fig.~\ref{fig:Y} in Sec.~\ref{sec:conclusion} summarizes the transverse and longitudinal gravitino abundance.

\subsubsection*{Abundance of inflatino}

Finally, we comment on the abundance of inflatino.
One should also be cautious about the notion of inflatino:
it is $\widetilde\phi$ if the SUSY breaking is dominated by $z$, \textit{i.e.,} $H \lesssim m_{3/2}^0$
while it is absorbed by the longitudinal gravitino at $H \gtrsim m_{3/2}^0$.
Thus, the heavy mass eigenstate of $(\psiLc, v_\perp)$ system eventually becomes the inflatino no matter when it is produced.

To estimate the inflatino abundance, we note that the heavy mass eigenvalue is given by
\begin{align}
	m_{\rm heavy} \simeq m_\phi + 2 \widehat{m}_{3/2},
\end{align}
where the second term is the oscillating part, see App.~\ref{sec:mass}.
In contrast to $m_{\rm light}$, there appears a term of $\mathcal O(\widehat{m}_{3/2})$ in $m_{\rm heavy}$, which can make the production efficient.
Just after inflation, $|\widehat{m}_{3/2}|$ is of the same order of $m_\phi$ if $\phi_i \sim \Mpl$, 
hence the production is kinematically accessible around $H \sim H_{\rm inf}$ and the production becomes 
less and less efficient after that due both to the kinematical suppression and the decrease of the oscillation amplitude of $\widehat{m}_{3/2}$.
Thus, the production of heavy mass eigenstate is dominated at $H \sim H_{\rm inf}$ and the resultant abundance is
of the same order of the transverse gravitino:
\begin{align}
	\frac{n_{\widetilde\phi}}{s} \simeq 
	\frac{27\mathcal C}{16\pi}\frac{H_{\rm inf} T_{\rm R}}{\Mpl^2}
	\simeq 9\times10^{-15} 
	\mathcal C
	\left( \frac{H_{\rm inf}}{10^{13}\,{\rm GeV}} \right)
	\left( \frac{T_{\rm R}}{10^{10}\,{\rm GeV}} \right),
\end{align}
where we have used $\widetilde{m} \simeq 2 \widehat{m}_{3/2} \simeq 3 H_{\text{inf}} \simeq \sqrt{3} m_{\phi}$ with $\widetilde{m}$ being the amplitude of the oscillating mass defined in App.~\ref{sec:fermion}.

The fate of the inflatino significantly depends on its interactions unspecified here.
Since we need to reheat the universe, there must be interactions of the inflaton superfield with the SUSY standard model (SSM) sector,
which automatically introduce inflatino-SSM interactions. 
It may also decay into gravitino if kinematically allowed~\cite{Nilles:2001my},
although such a contribution is smaller than that of the pre-existing transverse gravitino (\ref{trans_single}).

\subsection{Inflation model with a higher power potential}  \label{sec:higher_single}

In this subsection, we briefly discuss the inflation model with higher power.
The K\"ahler and superpotential are assumed to be
\begin{align}
	&K = -\frac{1}{2}(\phi-\phi^\dagger)^2 + |z|^2 - \frac{|z|^4}{\Lambda^2}, \\
	&W= \frac{1}{n}\lambda \phi^n + \mu^2 z + W_0.  
\end{align}
All parameters are taken to be real and positive without loss of generality.
It should be noticed that for $n>2$, the presence of constant $W_0$ eventually leads to the spontaneous breaking of $\mathbb{Z}_2$ or $\mathbb{Z}_n$ 
so that $\phi$ obtains a finite VEV.
When discussing a theory of higher $n$ $(>2)$ hereafter, we implicitly assume that there is also a quadratic term $W \sim m_\phi\phi^2$
which is subdominant at the early stage of (p)reheating,
and the $\mathbb{Z}_2$ symmetry is maintained in a practical sense.
For consistency with the Planck observation, we need modifications on the inflaton potential in the large field region,
and it can also simultaneously affect the potential in the reheating stage.
Such effects may be taken into account by including higher powers in the K\"{a}hler or superpotential.

The theoretical construction is almost parallel to the case of $n=2$ discussed so far after replacing $m_\phi \to \lambda\phi^{n-2}$.
One of the significant differences is the background evolution.
Let us suppose that the inflaton field $\phi$ oscillates around the potential minimum $\phi=0$ under a potential $V \simeq \lambda^2\phi^{2(n-1)}$.
The inflaton amplitude and the Hubble parameter decrease as
\begin{align}
	\phi_{\rm amp} \propto a^{-3/n},~~~~H^2 \propto a^{-6(n-1)/n}.  \label{H_higher}
\end{align}
This changes the abundance of gravitino, in particular the transverse one.

For example, for the quartic inflaton potential $n=3$, such as the Higgs-inflation like model,
it is much more abundant than the case of $n=2$:
\begin{align}
	\frac{n_{3/2}^{(t)}}{s} & \simeq 
	\frac{\mathcal C}{64\pi}\left( \frac{90}{\pi^2 g_*} \right)^{1/4}\left(\frac{H_{\rm inf}}{\Mpl}\right)^{3/2}
	\simeq 2\times10^{-11} \mathcal C\left( \frac{H_{\rm inf}}{10^{13}\,{\rm GeV}} \right)^{3/2},
	\label{Y_single2} \\
	\frac{n_{3/2}^{(\ell)}}{s} & \simeq 
	\frac{\mathcal C}{64\pi}\left( \frac{90}{\pi^2 g_*} \right)^{1/4}\left(\frac{m_{3/2}^0}{\Mpl}\right)^{3/2}
	\simeq 6\times10^{-22} \mathcal C\left( \frac{m_{3/2}^0}{10^{6}\,{\rm GeV}} \right)^{3/2},
\end{align}
independently of the reheating temperature.
Here we take $\mathcal C = 1$ as a reference.\footnote{
	For $n = 3$, the gravitino mass term may contain frequencies of $\Omega = m_\phi, 3 m_\phi$ 
	with $ m_\phi = \lambda \phi $. 
	We do not specify which process is dominant, 
	rather simply evaluate the rate in the case of $\Omega = m_\phi$
	as an order of magnitude estimation.
}
Therefore in this case the transverse gravitino is problematic even if the reheating temperature is very low.
The abundance of longitudinal gravitino does not change much compared with the case of $n=2$ after replacing $m_\phi \to m_\phi^{\rm eff}=\lambda\phi_{\rm amp}^{n-2}$.
This is because the transverse gravitino production is dominated at $H=H_{\rm inf}$ just after inflation
while the longitudinal one is dominated at $H \sim m_{3/2}^0$.
Thus the transverse gravitino abundance is much more sensitive to the background evolution than the longitudinal one.
See the dashed line in the left panel of Fig.~\ref{fig:Y} in Sec.~\ref{sec:conclusion} for the transverse and longitudinal gravitino abundance for $n=3$.

Another comment is that the inflaton ``mass'' $\sim \lambda\phi^{n-2}$ itself is a rapidly oscillating function for $n>2$.
It does not much affect the production of the light mass eigenstate of $(\psi^\ell, v_\perp)$,
but it can significantly change the production of the heavy mass eigenstate, which eventually becomes the inflatino (see Fig.~\ref{fig:1}).
Therefore, the final inflatino abundance is expected to be sensitive to the power $n$.\footnote{
	The enhancement of gravitino abundance for higher $n$ was already mentioned in Ref.~\cite{Kallosh:1999jj},
	although actually it should be regarded as the (final) inflatino abundance rather than the gravitino.
}
Although the inflatino seems to be less harmful than the gravitino,
precise discussion requires the information of interactions between the inflaton sector and the SSM sector for the reheating.

\subsection{Comment on the linear term in K\"ahler potential}   \label{sec:comment}

Here we comment on the case where the inflaton does not possess the $\mathbb{Z}_2$ symmetry and there is a linear term in the K\"ahler potential
\begin{align}
	K = ic(\phi-\phi^\dagger).   \label{linear}
\end{align}
As seen below, the existence of this term drastically changes the picture of the longitudinal gravitino production.

For the moment, we focus on the case of $n=2$.
First we should notice that there arises a mixing between $\phi$ and $z$ if there is a linear term in the K\"ahler potential as in Eq.~(\ref{linear}).
The mixing comes from
\begin{align}
	V \supset (D_\phi W)\overline{(D_\phi W)} \supset \frac{K_\phi W}{\Mpl^2}m_\phi \phi^* +{\rm h.c.}
	\simeq \frac{\sqrt{3}ic m_{3/2}m_\phi}{\Mpl} z \phi^* + {\rm h.c.}
\end{align}
The mixing angle is
\begin{align}
	\theta_{\phi z} \sim \begin{cases}
		\displaystyle\frac{\sqrt{3}c m_{3/2} m_\phi}{m_z^2 \Mpl}  & {\rm for}~~m_\phi < m_z\\ ~&~\\
		\displaystyle\frac{\sqrt{3}c m_{3/2}}{m_\phi \Mpl} & {\rm for}~~m_\phi > m_z
	\end{cases}.
\end{align}
As shown in App.~\ref{sec:dyn}, the oscillation of $z$ is induced by the $\phi$ oscillation, 
and its amplitude is given by $z_{\rm amp} \sim \theta_{\phi z} \phi_{\rm amp}$.
This induced $z$ oscillation efficiently produces the longitudinal gravitino through the $\delta z$ term in Eq.~(\ref{zLL}) for $H \lesssim m_{3/2}$.
The longitudinal gravitino production rate is thus given by
\begin{align}
	\dot n^{(\ell)}_{3/2} \simeq \frac{2\rho_z}{m_z}\Gamma(z\to\psiL\psiL) \simeq \frac{2\rho_\phi}{m_\phi} \frac{m_\phi}{m_z}\theta_{\phi z}^2\Gamma(z\to\psiL\psiL),
\end{align}
where
\begin{align}
	\Gamma(z\to\psiL\psiL) \simeq \frac{1}{96\pi} \frac{m_z^5}{(m^0_{3/2})^2\Mpl^2},
\end{align}
denotes the Polonyi decay width into the longitudinal gravitino pair.

\begin{align}
	\Gamma(\phi\to \psi^\ell \psi^\ell) \simeq \frac{m_\phi}{m_z}\theta_{\phi z}^2 \Gamma(z\to\psi^\ell\psi^\ell)\simeq
	\begin{cases}
		\displaystyle\frac{c^2 m_\phi^3}{32\pi \Mpl^4}  & {\rm for}~~m_\phi < m_z\\ ~&~\\
		\displaystyle\frac{c^2 m_\phi^3}{32\pi \Mpl^4} \left(\frac{m_z}{m_\phi}\right)^4 & {\rm for}~~m_\phi > m_z
	\end{cases}.
	\label{Gamma_pert}
\end{align}
This is consistent with Refs.~\cite{Endo:2007sz,Nakayama:2012hy}.
In contrast to the case of the previous subsection, this process is interpreted as ``decay'' of $\phi$ because the relevant interaction involves a single power of $\delta z$ and the mixing angle is constant. In other words, $\psiL$ ``feels'' the oscillation of $\phi$ linearly.
Thus, this process is more and more effective at later time.
The longitudinal mode produced during $H \lesssim m_{3/2}$ will eventually become the present gravitino.
The resulting abundance is estimated as
\begin{align}
	\frac{n_{3/2}^{(\ell)}}{s} & \simeq \left(\frac{2\Gamma(\phi\to \psiL \psiL)}{H}\right)_{H=\Gamma_{\rm inf}} \frac{3 T_{\rm R}}{4m_\phi}  
	\simeq \frac{3 c^2 m_{\phi}^2 }{64 \pi \Mpl^3 T_{\text{R}}} \left( \frac{90}{\pi^2 g_*} \right)^{1/2} \nonumber \\
	& \simeq 1\times 10^{-5}\left( \frac{c}{\Mpl} \right)^2
	\left( \frac{m_\phi}{10^{13}\,{\rm GeV}} \right)^2\left( \frac{10^{10}\,{\rm GeV}}{T_{\rm R}} \right),
\end{align}
where $\Gamma_{\rm inf}$ denotes the total decay width of the inflaton.
Cosmological consequences of this kind of scenario is extensively studied in Refs.~\cite{Endo:2007sz,Nakayama:2012hy}.
This abundance is quite huge and it is one of the motivations to consider the $\mathbb{Z}_2$-symmetric model.\footnote{
If the inflaton and Polonyi sectors are effectively sequestered, the enhanced gravitino production can be avoided even when the inflation sector breaks the $\mathbb{Z}_2$ symmetry.  One of the simplest ways is to move the linear term to the superpotential by the K\"{a}hler-Weyl transformation before coupling it to the Polonyi sector.
} 
In the $\mathbb{Z}_2$-symmetric case $c=0$, the decay process does not exist, because there is no mixing between $z$ and $\phi$.\footnote{
	In Ref.~\cite{Dine:2006ii}, it was pointed out that the gravitino does not couple to the heavy mass eigenstate of $(\phi, z)$ system
	in the limit $m_z \ll m_\phi$. This corresponds to $(m_z/m_\phi)^4$ suppression in (\ref{Gamma_pert})~\cite{Endo:2006tf}.
	In the $\mathbb{Z}_2$-symmetric case, this statement is trivial because $m_{3/2}$ does not explicitly contain $\phi$ but only $z$ at the linear level.
}

A similar result holds for general $n (> 2)$ after replacing $m_\phi$ with the effective inflaton mass $m_\phi^{\rm eff}\sim \lambda \phi_{\rm amp}^{n-2}$.
In this case, however, the partial ``annihilation'' rate $\Gamma(\phi\phi\to \psiL \psiL)$ becomes smaller for late time and the gravitino production is suppressed compared with the case of $n=2$.

\section{Gravitino production in multi-superfield inflation} \label{sec:multi}

We consider a more realistic three-field case: the inflaton $\phi$, the stabilizer $X$ and the Polonyi field $z$.
The K\"ahler and superpotential are given by
\begin{align}
	&K = - \frac{1}{2}(\phi-\phi^\dagger)^2 + |X|^2 + |z|^2 - \frac{|z|^4}{\Lambda^2}, \\
	&W= \lambda X\phi^n + \mu^2 z + W_0.
\end{align}
This model has an approximate shift symmetry $\phi \to \phi + c$ with $c$ being a real constant, 
and the inflaton is identified with $\sqrt{2}\,{\rm Re}(\phi)$~\cite{Kallosh:2010ug, Kallosh:2010xz}.
Without loss of generality, we can take all the coupling  constants real and positive.
Simple power-law chaotic inflation models are not favored according to the results from the Planck satellite~\cite{Ade:2015lrj},
and hence we need modifications on the models to be consistent with observation~\cite{Kawasaki:2000yn,Kallosh:2010ug,Kallosh:2010xz,Ferrara:2010in,Nakayama:2010kt,Nakayama:2013txa,Kallosh:2013hoa,Kallosh:2013yoa,Galante:2014ifa}.
Again, we are interested in the sub-Planckian regime and these modifications on the potential in the large field region do not much affect the
following discussion.

From the viewpoint of gravitino production, the significant difference from the single-field case is that 
the stabilizer field $X$ is close to zero during inflation.\footnote{
	Precisely speaking, we need $K \sim -|X|^4/\Mpl^2$ to give $X$ the Hubble mass and stabilize $X\simeq 0$ during inflation,
	and also it slightly deviates from the origin: $X\sim (m_{3/2}^0/m_\phi) \phi$.
	However, it does not affect the following discussion.
}
It suppresses the oscillating gravitino mass $m_{3/2} \sim \lambda X\phi^n/\Mpl^2$, leading to a suppressed gravitino production rate.

\subsection{Dynamics}

In contrast to the single-field model, $m_{3/2} \simeq m_{3/2}^0$ just after inflation since $X$ is stabilized at $|X| \ll |\phi|$.
However, $|X|$ does not remain to be very small because of the mixing between $\phi$ and $X$:
\begin{align}
	V \supset m_{3/2}^0\lambda \phi^{n-1}\left[ (n-2)\phi - n \phi^* \right] X +{\rm h.c.}
\end{align}
Since we are assuming that all parameters in the model are real, the reality of $\phi$ and $X$ is also maintained.
Thus the potential around the minimum is given by
\begin{align}
	V = (\phi~X)
  \begin{pmatrix}
  	(\lambda \phi^{n-1})^2 & -2m_{3/2}^0 (\lambda \phi^{n-1}) \\
	-2m_{3/2}^0 (\lambda \phi^{n-1}) & n^2 (\lambda \phi^{n-1})^2
  \end{pmatrix}
   \begin{pmatrix}
  	\phi  \\
	X
  \end{pmatrix}.
\end{align}
According to App.~\ref{sec:dyn}, the $\phi$ oscillation induces the $X$ oscillation due to the mixing term.

For $n\neq 1$, masses of $\phi$ and $X$ are not degenerate and the amplitude of induced $X$ oscillation is about
$X_{\rm amp} \sim (m_{3/2}^0/m_\phi) \phi_{\rm amp}$, where $m_\phi \equiv \lambda \phi_{\rm amp}^{n-1}$ denotes the effective mass of the inflaton.
It means that $\phi$ and $X$ are maximally mixed with each other at $m_\phi \simeq m_{3/2}^0$, which happens at $H \ll m_{3/2}^0$.
Thus we have
\begin{align}
	m_{3/2} \simeq \begin{cases}
		\displaystyle\frac{m_{3/2}^0 \phi^2}{\Mpl^2}+ m_{3/2}^0 &{\rm for}~~m_\phi > m_{3/2}^0 \\
		\displaystyle\frac{m_\phi \phi^2}{\Mpl^2} + m_{3/2}^0 &{\rm for}~~m_\phi < m_{3/2}^0 
	\end{cases}.
	\label{mg_multi_n}
\end{align}
The overall magnitude remains to be about $m_{3/2}^0$ throughout the whole history of the universe.
The oscillating part is the first term.\footnote{
	Even if the term with $n>1$ is dominant at the early stage, eventually the term with $n=1$ becomes dominant.
	Otherwise, $\phi$ and $X$ would be massless in the present vacuum.
	The bare mass term $n=1$ can change the dynamics described above.
}

For $n=1$, on the other hand, masses of $\phi$ and $X$ are almost degenerate.
As shown in App.~\ref{sec:dyn}, for $H \gtrsim m_{3/2}^0$, the amplitude is given by\footnote{
	The Hubble mass term like $\sim H^2 |X|^2$ does not change the discussion. 
}
\begin{align}
	X_{\rm amp} \sim \phi_i \left(\frac{a_i}{a(t)}\right)^{3/2} m_{3/2}^0 t \sim \phi_i \left(\frac{m_{3/2}^0}{H_{\rm inf}}\right) \sim 
	\frac{m_{3/2}^0}{H} \phi_{\rm amp}.
\end{align}
Notably, this is almost constant until $H \sim m_{3/2}^0$.
After that, $\phi \sim X$ have the same order of oscillation amplitudes, meaning the maximal mixing of $\phi$ and $X$.
This implies that
\begin{align}
	m_{3/2} \simeq \begin{cases}
		\displaystyle\frac{m_\phi \phi^2}{\Mpl^2}\frac{m_{3/2}^0}{H} + m_{3/2}^0 &{\rm for}~~H > m_{3/2}^0 \\
		\displaystyle\frac{m_\phi \phi^2}{\Mpl^2} + m_{3/2}^0 &{\rm for}~~H < m_{3/2}^0 
	\end{cases}.
	\label{mg_multi_1}
\end{align}
The overall magnitude is about $m_{3/2}^0$ throughout the whole history of the universe.
The first term is the oscillating part, which is suppressed by $m_{3/2}^0/H$ at $H > m_{3/2}^0$ compared with the single-field case.

The dynamics of Polonyi field $z$ is similar to the single-superfield inflation case studied in Sec.~\ref{sec:dyn_single}, so we do not repeat the analysis here.

Based on this picture, below we estimate the gravitino abundance for $n=1$ in which $W=m_\phi X\phi$.
The case of higher $n$ is discussed in Sec.~\ref{sec:higher_multi}.

\begin{figure}[t]
\begin{center}
\includegraphics[scale=0.8]{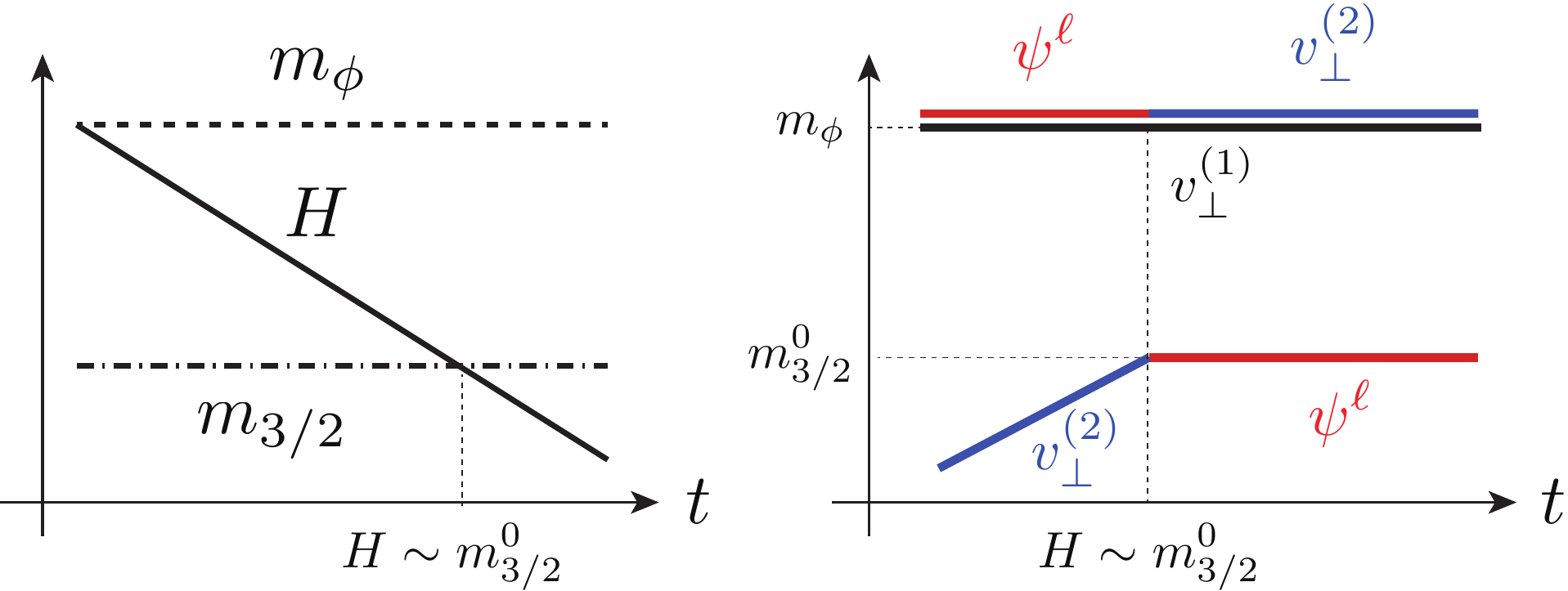}
\end{center}
\caption {\small
Same as Fig.~\ref{fig:1} but for multi-superfield inflation models.
}
\label{fig:2}
\end{figure}

\subsection{Transverse component}  \label{sec:trans_multi}

Let us estimate the transverse gravitino production rate for $n=1$.
As already shown, it is solely determined by the field-dependence of the gravitino mass $m_{3/2}$.
According to Eq.~(\ref{mg_multi_1}), it is suppressed by 
the small factor $m_{3/2}^0/H$ compared with the single-field case studied in Sec.~\ref{sec:trans_single}.
Thus we obtain the effective ``annihilation'' rate of the inflaton into the transverse gravitino pair as 
\begin{align}
	\Gamma(\phi\phi\to\psiT\psiT) \simeq 
	\begin{cases}
	\displaystyle
	\frac{\mathcal C}{4\pi} \frac{\phi_{\rm amp}^2}{\Mpl^2} \frac{m_\phi^3}{\Mpl^2}
	\left(\frac{m_{3/2}^0}{H}\right)^2
	\simeq \frac{3\mathcal C}{4\pi}\frac{(m_{3/2}^0)^2 m_\phi}{\Mpl^2} & {\rm for}~~H>m_{3/2}^0\\[1em]
	\displaystyle
	 \frac{\mathcal C}{4\pi} \frac{\phi_{\rm amp}^2}{\Mpl^2} \frac{m_\phi^3}{\Mpl^2} 
	 \simeq \frac{3\mathcal C}{4\pi}\frac{H^2 m_\phi}{\Mpl^2}
	 & {\rm for}~~H<m_{3/2}^0
	 \end{cases}.
\end{align}
In contrast to the single-superfield inflation case, the production rate is time-independent at the early epoch,
hence the transverse gravitino production is maximized at $H \sim m_{3/2}^0$.
Thus the abundance of transverse gravitino is found to be
\begin{align}
	\frac{n_{3/2}^{(t)}}{s} 
	&\simeq \left(\frac{\Gamma(\phi\phi\to \psiT\psiT)}{H}\right)_{H=m_{3/2}^0} \frac{3T_{\rm R}}{4m_\phi}
	= \frac{9\mathcal C}{16\pi} \frac{m_{3/2}^0 T_{\rm R}}{\Mpl^2} \nonumber\\
	& \simeq 3 \times 10^{-22}\,
	\mathcal C
	\left( \frac{m_{3/2}^0}{10^6\,{\rm GeV}} \right)
	\left( \frac{T_{\rm R}}{10^{10}\,{\rm GeV}} \right).
\end{align}
This is similar to the abundance of longitudinal gravitino in the single-superfield inflation case (\ref{Y_single})
and much smaller than the gravitino abundance from thermal production.
Thus it does not give significant effects on cosmology.

\subsection{Longitudinal component} \label{sec:long_multi}

Next, let us discuss the longitudinal gravitino production.
The discussion is parallel to the previous single-superfield inflation case, but one care is needed
since it is $|F_X| \sim |\dot\phi| \sim H\Mpl$ that dominantly breaks SUSY at $H > m_{3/2}^0$.
It may be useful to rewrite the K\"ahler and superpotential as
\begin{align}
	&K = |\Phi_+|^2+|\Phi_-|^2 -\frac{1}{4}\left[\left(\Phi_++\Phi_-\right)^2+{\rm h.c.} \right]  + |z|^2 - \frac{|z|^4}{\Lambda^2}, \\
	&W=\frac{1}{2}m_\phi\left(\Phi_+^2-\Phi_-^2\right) + \mu^2 z + W_0,
\end{align}
where
\begin{align}
	\Phi_+ \equiv \frac{1}{\sqrt{2}}(\phi+X),~~~~~~\Phi_- \equiv \frac{1}{\sqrt{2}}(\phi-X). \label{light-cone_basis}
\end{align}
In this basis, we can follow the same method as that in Sec.~\ref{sec:multigrav} to derive the Lagrangian of the gravitino-fermion system.
The goldstino is given by
\begin{align}
	v = \frac{1}{\sqrt{\rho_{\rm SB}}}\left[ \sqrt{\rho_{\rm SB}^+}v_+ +  \sqrt{\rho_{\rm SB}^-}v_-
	+ \sqrt{\rho_{\rm SB}^z} v_z   \right] 
	\equiv \alpha_+ v_+ + \alpha_-v_- + \alpha_z v_z,
\end{align}
where
\begin{align}
	&v_+ \equiv \frac{1}{\sqrt{\rho_{\rm SB}^+}}\left(-F_{\Phi_+} + \dot\Phi_+\gamma^0\right)\widetilde\Phi_+ \equiv e^{\gamma^0\theta_+}\widetilde\Phi_+,
	~~~\rho_{\rm SB}^+ \equiv |\dot \Phi_+|^2 + |F_{\Phi_+}|^2,\\
	&v_- \equiv \frac{1}{\sqrt{\rho_{\rm SB}^-}}\left(-F_{\Phi_-}+ \dot \Phi_- \gamma^0\right) \widetilde\Phi_- \equiv e^{\gamma^0\theta_-}\widetilde\Phi_-,
	~~~\rho_{\rm SB}^- \equiv |\dot \Phi_-|^2 + |F_{\Phi_-}|^2,\\
	&v_z \equiv  \frac{1}{\sqrt{\rho_{\rm SB}^z}}\left(-F_z + \dot z\gamma^0\right)\widetilde z \equiv e^{\gamma^0\theta_z}\widetilde z,
	~~~\rho_{\rm SB}^z \equiv |\dot z|^2 + |F_z|^2.
\end{align}
We can define the $3\times 3$ orthogonal matrix $O$ that transforms the original basis into $(v,v_\perp)$ basis: 
\textit{e.g.},
\begin{align}
	\begin{pmatrix}
		v \\ v_\perp^{(1)} \\ v_\perp^{(2)}
	\end{pmatrix}
	= O^T
	\begin{pmatrix}
	e^{\gamma^0\theta_+}\widetilde\Phi_+ \\ e^{\gamma^0\theta_-}\widetilde\Phi_- \\ e^{\gamma^0\theta_z}\widetilde z
	\end{pmatrix},
	~~~~~~
	O^T = 
	\begin{pmatrix}
		\alpha_+ ~\alpha_- ~\alpha_z \\
                  \widetilde O^T		
	\end{pmatrix}
	\equiv
	\begin{pmatrix}
		s_1c_2 & c_1c_2 & s_2 \\
		c_1 & -s_1 & 0 \\
		-s_1s_2& -c_1 s_2 & c_2
	\end{pmatrix},
\end{align}
where $s_1^2+c_1^2=s_2^2+c_2^2=1$ and $\widetilde O$ is the $3\times 2$ matrix.
It is easily checked that we have 
 $s_2\simeq 0$ and $v \simeq s_1 e^{\gamma^0\theta_+}\widetilde \Phi_+ + c_1 e^{\gamma^0\theta_-}\widetilde \Phi_-$
 and $v_\perp^{(2)} \sim e^{\gamma^0\theta_z} \widetilde z$ for $H\gg m_{3/2}^0$,
while $s_2\simeq 1$ and $v \simeq e^{\gamma^0\theta_z}\widetilde z$
and $v_\perp^{(2)} \sim -s_1 e^{\gamma^0\theta_+}\widetilde \Phi_+ - c_1 e^{\gamma^0\theta_-}\widetilde \Phi_-$
in the opposite limit $H\ll m_{3/2}^0$.

Following the same procedure as in Sec.~\ref{sec:multigrav}, we obtain the
Lagrangian of the longitudinal gravitino and fermion system as
\begin{align}
	\mathcal L_{f} = -\frac{1}{2}\begin{pmatrix}
	\overline{{\psiLc}'} & \overline{{v_\perp^{(1)}}'} & \overline{{v_\perp^{(2)}}'} 
	\end{pmatrix}
	\left[
	\gamma^0\partial_0 + i\vec\gamma\cdot\vec k
	+a\mathcal M
	\right]
	\begin{pmatrix}
	{\psiLc}' \\ {v_\perp^{(1)}}' \\ {v_\perp^{(2)}}' 
	\end{pmatrix},
\end{align}
where $\mathcal M$ denotes the mass matrix defined in Eq.~(\ref{mass_multi}), whose determinant is
\begin{align}
	\det\mathcal M \simeq m_{3/2}\, \det \left(\widetilde{O}^T \widehat m_f \widetilde{O} \right) \simeq \alpha_z^2 m_{3/2} m_\phi^2.
\end{align}
Although a full expression of $\mathcal M$ is lengthy, we can deduce the mass eigenvalues as
\begin{align}
	\begin{pmatrix}
	 m_{\rm heavy}^{+}, & m_{\rm heavy}^{-}, &
	 m_{\rm light}
	\end{pmatrix}
	\simeq 
	\begin{cases}
	\begin{pmatrix}
	 m_{\phi}, & -m_\phi, &
	 -m_{3/2}(m_{3/2}^0/H)^2
	\end{pmatrix}
	&{\rm for}~~H\gtrsim m_{3/2}^0
	\\
	\begin{pmatrix}
	 m_{\phi},& -m_\phi, &
	 -m_{3/2}
	\end{pmatrix}
	&{\rm for}~~H\lesssim m_{3/2}^0
	\end{cases}.
	\label{mHmHmL}
\end{align}
Therefore the expression for $m_{\rm light}$ is similar to the single-superfield inflation case.
The lightest mass eigenstate, which should be mostly composed of $\widetilde z$, 
is $v_\perp^{(2)}$ for $H \gg m_{3/2}^0$ and $\psi_c^\ell$ for $H\ll m_{3/2}^0$.
A more detailed explanation for the mass eigenvalues is given in App.~\ref{sec:mass}. 
The right panel of Fig.~\ref{fig:2} shows the schematic picture of the time evolution of the absolute values of the mass eigenvalues.

\subsubsection*{Abundance of longitudinal gravitino}

The lightest mass eigenstate eventually becomes the present-day longitudinal gravitino.
Hence, we here would like to estimate the production rate of the lightest mass eigenstate in the universe. 
Recalling that $m_{3/2}$ in the inflation model with a stabilizer field is given by Eq.~(\ref{mg_multi_1}), 
and repeating the same analysis for the production rate as that in Sec.~\ref{sec:long_single}, we find that an upper bound on the longitudinal gravitino production rate as
\begin{align}
	\Gamma(\phi\phi\to v_\perp^{(2)} v_\perp^{(2)}) \lesssim \frac{\mathcal C}{4\pi}\left( \frac{m_{3/2}^0}{H} \right)^6 \frac{\phi_{\rm amp}^2}{\Mpl^2} \frac{m_\phi^3}{\Mpl^2}
	\simeq \frac{3\mathcal C}{4\pi}\left( \frac{m_{3/2}^0}{H} \right)^4\frac{(m_{3/2}^0)^2 m_\phi}{\Mpl^2},
\end{align}
for $H > m_{3/2}^0$, and
\begin{align}
	\Gamma(\phi\phi\to\psiL\psiL) \lesssim \frac{\mathcal C}{4\pi} \frac{\phi_{\rm amp}^2}{\Mpl^2} \frac{m_\phi^3}{\Mpl^2}
	\simeq \frac{3\mathcal C}{4\pi}\frac{H^2 m_\phi}{\Mpl^2},
\end{align}
for $H < m_{3/2}^0$.
We again find that the resulting abundance is dominated by those created around $H\sim m_{3/2}^0$, with an amount of
\begin{align}
	\frac{n_{3/2}^{(\ell)}}{s} &\lesssim \left(\frac{ \Gamma(\phi\phi\to \psiL\psiL)}{H}\right)_{H=m_{3/2}^0} \frac{3T_{\rm R}}{4m_\phi}
	\simeq \frac{9\mathcal C}{16\pi} \frac{m_{3/2}^0 T_{\rm R}}{\Mpl^2} \nonumber\\
	&\simeq 3\times 10^{-22}\,
	\mathcal C
	\left( \frac{m_{3/2}^0}{10^{6}\,{\rm GeV}} \right)
	\left( \frac{T_{\rm R}}{10^{10}\,{\rm GeV}} \right).
\end{align}
It is the same order of the abundance of the transverse gravitino and also comparable to the longitudinal gravitino abundance in the
case of single-superfield inflation model (\ref{Y_single}).
It is too small to give significant phenomenological effects.
As noted in the single-superfield infaltion case, the contribution from the induced $z$ oscillation to the longitudinal gravitino abundance
is at most comparable to this upper bound.
After all, in inflation models with a stabilizer field $X$, it is safe to neglect the nonthermal gravitino production after inflation.
The dotted line in the right panel of Fig.~\ref{fig:Y} in Sec.~\ref{sec:conclusion} summarizes the transverse and longitudinal gravitino abundance 
in models with a stabilizer field.

\subsubsection*{Abundance of inflatino and stabilizino}

There are two heavy mass eigenstates in the present model.
They are roughly $(\psi^\ell, v_\perp^{(1)})$ at the early epoch $(H>m_{3/2}^0)$
and $(v_\perp^{(1)}, v_\perp^{(2)})$ at the late epoch $(H < m_{3/2}^0)$.
They are finally regarded as inflatino/stabilizino fields, since both $v_\perp^{(1)}$ and $v_\perp^{(2)}$ are
composed of $\widetilde\phi$ and $\widetilde X$ at late epoch.
The heavy mass eigenvalues are given by
\begin{align}
	m_{\rm heavy}^{\pm} \simeq \pm m_\phi + 2 \alpha_{\pm}^2 \widehat{m}_{3/2}^{\pm},
\end{align}
and $\widehat{m}_{3/2}^{\pm}$ contains an oscillating term of $\mathcal O(H)$ (see App.~\ref{sec:mass}).
Thus the production rate of the heavy mass eigenstates at $H\gg m_{3/2}^0$ is similar to the case without the stabilizer field.
This process is accessible only just after inflation when $H\sim m_\phi$.
The production of heavy states is expected to be dominated at $H \sim H_{\rm inf}$, and the resulting inflatino/stabilizino abundance at late time is
estimated to be
\begin{align}
	\frac{n_{v_\perp^{(1)}}}{s} \simeq \frac{n_{v_\perp^{(2)}}}{s} \simeq  \frac{27\mathcal C}{16\pi} \frac{H_{\text{inf}} T_{\rm R}}{\Mpl^2}
	\simeq 9\times 10^{-15} \,\mathcal C
	\left( \frac{H_{\text{inf}}}{10^{13}\,{\rm GeV}} \right)^{-1}
	\left( \frac{T_{\rm R}}{10^{10}\,{\rm GeV}} \right),
\end{align}
which is similar to the inflatino abundance in the single-superfield inflation case.
Again we note that the fate of the inflatino and stabilizino depends on the inflaton-SSM interactions which is not specified here.
Generally the inflatino decay into the gravitino plus inflaton may be kinematically allowed depending on the 
soft SUSY breaking mass of the inflaton, but the branching fraction is expected to be small.
However, depending on the inflatino/stabilizino branching fractions into gravitino, this channel can be the dominant source of nonthermal gravitino.

\subsection{Inflation model with a higher power potential}  \label{sec:higher_multi}

So far we have focused on the case of quadratic inflaton potential $n=1$.
Now let us briefly discuss the case with a higher power $n >1$.
As described in Sec.~\ref{sec:higher_single}, there are mainly two differences from the quadratic case.  
One is the change of the background evolution: see Eq.~(\ref{H_higher}).
The other is that the inflaton mass itself becomes a rapidly oscillating function.
Although the full analysis is complicated, we can qualitatively discuss these effects.

First, 
the transverse gravitino abundance is dominated at $H \sim H_\text{inf}$, but it is suppressed by a factor $(m_{3/2}^0/m_{\phi})^2$ compared to the single-superfield case because of the suppression of the gravitino mass~\eqref{mg_multi_n}.
The production of the longitudinal component is dominated at $H \sim m_{3/2}^0$.
For example, for the quartic inflaton potential $n=2$, we obtain 
\begin{align}
	\frac{n_{3/2}^{(t)}}{s} 
	&\simeq 
	\frac{9 \mathcal C}{16\pi}\left( \frac{90}{\pi^2 g_*} \right)^{1/4}\left(\frac{(m_{3/2}^0)^2}{H_{\rm inf}^{1/2}\Mpl^{3/2}}\right)
	\simeq 7\times10^{-24} \mathcal C\left( \frac{H_{\rm inf}}{10^{13}\,{\rm GeV}} \right)^{-1/2}
	\prn{ \frac{m_{3/2}^0}{10^6 \, {\rm GeV}} }^2, \\
	\frac{n_{3/2}^{(\ell)}}{s}
	&\simeq \frac{9 \mathcal C}{16\pi}\left( \frac{90}{\pi^2 g_*} \right)^{1/4} \left( \frac{m_{3/2}^0}{\Mpl} \right)^{3/2}
	\simeq 2\times 10^{-20}\,\mathcal C
	\left( \frac{m_{3/2}^0}{10^{6}\,{\rm GeV}} \right)^{3/2},
\end{align}
where $\Gamma_\phi$ denotes the total decay width of the inflaton.
See the dashed line in the right panel of Fig.~\ref{fig:Y} in Sec.~\ref{sec:conclusion} for the transverse and longitudinal gravitino abundance for $n=2$.
The oscillation of inflaton mass itself also does not much affect the final gravitino abundance,
since the mass eigenvalue of the light state is determined by $m_{3/2}$, not the inflaton mass.
On the other hand, the inflatino abundance can be enhanced
since its production is dominated at the early epoch $H \sim H_{\rm inf}$, and the oscillation of the effective inflaton mass itself
directly contributes to the heavy mass eigenstates.

\section{Conclusions}\label{sec:conclusion}

\begin{figure}[t]
\begin{center}
\includegraphics[scale=1.3]{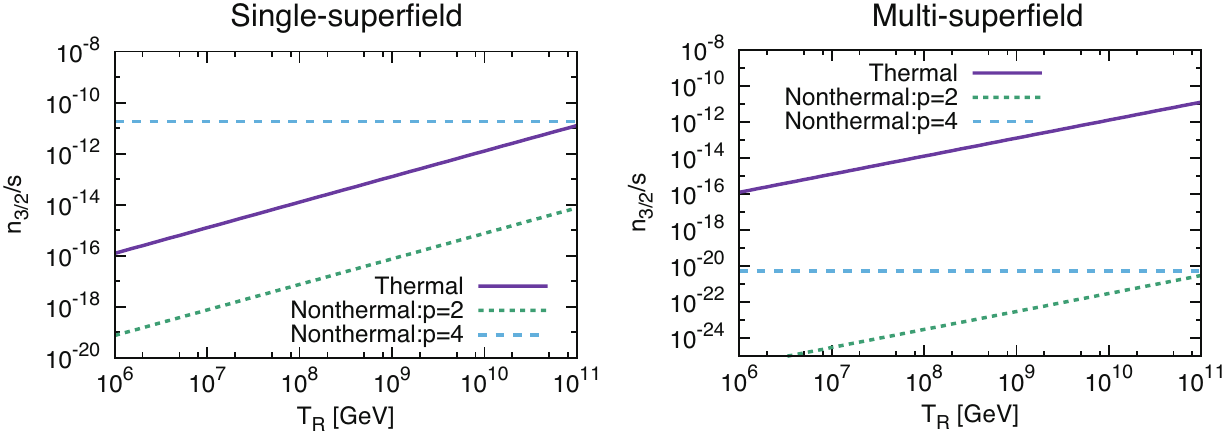}
\caption{\small
	Gravitino abundance $n_{3/2}/s$ as a function of reheating temperature $T_{\rm R}$ for the single-superfield inflation model without a stabilizer field (\textbf{left})
	and multi-superfield inflation model with a stabilizer field (\textbf{right}).
	The solid line shows a contribution from thermal production, while dotted (dashed) lines show nonthermally produced ones
	for the inflaton potential $V\propto \phi^p$ with $p=2$ ($p=4$).
}\label{fig:Y}
\end{center}
\end{figure}

We have studied the nonthermal gravitino production during (p)reheating paying particular attention to the case of $\mathbb{Z}_2$ symmetric large field inflation models.
The result crucially depends on inflation models.
In single-superfield inflation without a stabilizer field, production of the transverse gravitino is efficient and it can cause cosmological problems depending on the power law index of the inflaton potential.
The longitudinal gravitino production is safely neglected.
On the other hand, in multi-superfield inflation models with a stabilizer field, the transverse gravitino production is significantly suppressed
and nonthermal gravitino production plays no important role in cosmology.
Fig.~\ref{fig:Y} shows the gravitino abundance as a function of the reheating temperature $T_{\rm R}$ for the single-superfield inflation models (left)
and multi-superfield inflation models with a stabilizer field (right).
The solid line shows a contribution from thermal production~\cite{Bolz:2000fu,Pradler:2006qh,Rychkov:2007uq}, 
while dashed (dotted) lines show nonthermally produced ones
for the inflaton potential $V\propto \phi^p$ with $p=2$ ($p=4$).
Here we have taken $H_{\rm inf}=10^{13}\,$GeV (left) and $m_{3/2}^0=10^6$\,GeV (right).
If the inflaton potential changes from $\phi^4$ to $\phi^2$ at some point, the prediction lies between these two lines.
It is clearly seen that inflation models with a stabilizer predicts negligibly small nonthermal gravitino abundance.
Therefore inflation models with a stabilizer field is motivated not only from the viewpoint of model building,
but also from the requirement to avoid the nonthermal gravitino overproduction.
Note that in this plot we have ignored some other nonthermal gravitino production processes such as those from Polonyi/inflatino decay
since they are rather model-dependent~\cite{Nakayama:2012hy,Nilles:2001my}.
However, inclusion of them does not much affect this conclusion.

Some comments are in order.
In the most part we assumed the (nearly) minimal K\"ahler potential for the inflaton superfield for simplicity.
The production rate is significantly modified if there is a $\mathbb{Z}_2$-symmetric K\"ahler potential of the form
\begin{align}
	K \sim \frac{1}{\Mpl^2}|\phi|^2 zz +{\rm h.c.}~~~{\rm or}~~~ \frac{1}{\Mpl^2}X^\dagger\phi zz + {\rm h.c.},
\end{align}
for the single-superfield inflaton and multi-superfield inflaton case, respectively.
Since these terms directly give the large oscillating Polonyino mass like $m_{\tilde z} \sim m_\phi \phi^2/{\Mpl^2}$,
the longitudinal gravitino production rate is expected to be significantly enhanced to the same level as the transverse production rate,
if the coefficients of these terms are $\mathcal O(1)$.
Moreover, some inflation models, especially those constructed from the Jordan frame action,
have a nonminimal K\"ahler potential of the inflaton sector itself which can potentially induce violent phenomena~\cite{Ema:2016dny}.
Also we assumed that the inflaton is a gauge singlet:
for a gauge non-singlet inflaton, the structure becomes more complicated.
We will come back to these issues in future.

\section*{Acknowledgments}

This work was supported by the Grant-in-Aid for Scientific Research on Scientific Research A (No.26247042 [KN]),
Young Scientists B (No.26800121 [KN]) and Innovative Areas (No.26104009 [KN], No.15H05888 [KN]).
This work was supported by World Premier International Research Center Initiative (WPI Initiative), MEXT, Japan. 
This work is supported in part by National Research Foundation of Korea (NRF) Research Grant NRF-2015R1A2A1A05001869.
The work of Y.E. and K.M. was supported in part by JSPS Research Fellowships for Young Scientists.
The work of Y.E. was also supported in part by the Program for Leading Graduate Schools, MEXT, Japan.
A part of the work of T.T. was done when he belonged to APCTP.

\appendix
\section{Notations and conventions} \label{sec:notation}
In this appendix, we summarize the notations and conventions used in this paper.
We follow those of the textbook~\cite{Freedman:2012zz}.

\subsubsection*{Gamma matrix}
Here we summarize the notations and conventions related to the gamma matrices.
We take the metric to be ``almost plus'', \textit{i.e.,} $\eta_{\mu\nu} = \mathrm{diag}(-1, +1, +1, +1)$ 
for the flat space, and similar way for the FLRW one.
Then, the Clifford algebra is defined as
\begin{align}
	\left[\widehat{\gamma}^{\mu}, \widehat{\gamma}^{\nu}\right]_+ = 2g^{\mu\nu},
	\label{eq:Clifford}
\end{align}
where $\left[A, B\right]_+ = AB + BA$, and the hats denote the quantities 
in the curved space-time as it is also noted in the main text.
We express the gamma matrices in the flat space without the hats.
Through the vierbein $e_\mu^a$, they are related:
\begin{align}
	\widehat{\gamma}_{\mu} = e_\mu^a \gamma_a.
\end{align}
Here the vierbein is defined as
\begin{align}
	g_{\mu\nu} = \eta_{ab}e_{\mu}^{a}e_{\nu}^{b}.
\end{align}
The sign convention of Eq.~\eqref{eq:Clifford} as well as that of the metric
determine the (anti-)hermiticity of the gamma matrices.
In our convention, $\gamma^0$ is anti-hermitian, while $\vec \gamma$ is hermitian:
\begin{align}
	\left(\gamma^0\right)^\dagger = -\gamma^0, 
	~~ \left(\vec \gamma \right)^\dagger = \vec \gamma.
\end{align}
We define the short hand notations for the product of the gamma matrices as
\begin{align}
	\widehat{\gamma}_{\mu\nu} &\equiv \frac{1}{2}\left[\widehat{\gamma}_\mu, \widehat{\gamma}_\nu \right]_{-}, \\
	\widehat{\gamma}_{\mu\nu\rho} &\equiv \frac{1}{3}\left( \widehat{\gamma}_\mu \widehat{\gamma}_{\nu\rho}
	+ \widehat{\gamma}_\nu \widehat{\gamma}_{\rho\mu} + \widehat{\gamma}_\rho \widehat{\gamma}_{\mu\nu} \right),
\end{align}
where $[A, B]_- = AB - BA$.
Note that they are anti-symmetric with respect to the indices.
We also define the hermitian matrix $\gamma_*$ as
\begin{align}
	\gamma_* \equiv i \gamma_0 \gamma_1 \gamma_2 \gamma_3,
\end{align}
where we have used the gamma matrices in the flat space, not in the curved space.
In terms of it, the projection operators are defined as
\begin{align}
	P_L \equiv \frac{1 + \gamma_*}{2}, ~~ P_R \equiv \frac{1 - \gamma_*}{2}.
\end{align}

\subsubsection*{Dirac/Majorana conjugation}
The Dirac conjugation of a fermion $\psi$ is defined as
\begin{align}
	\overline{\psi} \equiv i\psi^\dagger \gamma^0.
\end{align}
In order to define the Majorana conjugation, we need the charge conjugation matrix $C$.
It is a unitary matrix that satisfies
\begin{align}
	\gamma^{\mu}{}^T = -C\gamma^\mu C^{-1}.
\end{align}
With this matrix, the Majorana conjugation is defined as
\begin{align}
	\overline{\psi}^C \equiv \psi^T C.
\end{align}
For the Majorana fermion $\lambda$ that satisfies
\begin{align}
	\lambda = -C^{-1}\overline{\lambda}^T,
	\label{eq:Majorana_cond}
\end{align}
the Dirac and Majorana conjugations are equivalent.\footnote{
	The sign in front of Eq.~\eqref{eq:Majorana_cond}
	is negative since we follow the convention of Ref.~\cite{Freedman:2012zz}.
} The explicit form of $C$ in terms of $\gamma$
depends on the representation of the gamma matrices.

\subsubsection*{Curvature}
First we define the spin connection as
\begin{align}
	\omega_{\mu}{}^{ab} = 2e^{\nu[a}\partial_{[\mu}e_{\nu]}{}^{b]}
	- e^{\nu[a}e^{b]\sigma}e_{\mu c}\partial_\nu e_{\sigma}{}^{c},
\end{align}
where we have neglected the torsion since it contributes only to the four Fermi interaction.
Here [...] denotes the anti-symmetrization with respect to those indices.
The normalization of the anti-symmetrization of $n$ indices includes the factor of $1/n!$.
In terms of it, the Riemann tensor is defined as
\begin{align}
	R_{\mu\nu}{}^{ab} \equiv 2 \partial_{[\mu}\omega_{\nu]}{}^{ab} 
	+ 2\omega_{[\mu}{}^{ac}\omega_{\nu]c}{}^{b}.
\end{align}
The Ricci tensor is defined as
\begin{align}
	R_{\mu\nu} \equiv R_{\mu\rho}{}^{ab}e_{a\nu}e_{b}{}^{\rho}.
\end{align}
Finally we define the Ricci scalar as
\begin{align}
	R \equiv g_{\mu\nu}R^{\mu\nu}.
\end{align}

\section{Fermion production in the background field method}  \label{sec:fermion}

In this appendix, we discuss production of a Majorana fermion due to a time-dependent mass term:
\begin{align}
	S = -\frac{1}{2}\int \dd^4 x\, \overline \psi \com{\slashed{\der}  - m (t)} \psi.
\end{align}
Though the time dependence of mass term typically arises from a coherently oscillating scalar field in the cosmological context,
here we do not specify the origin of $m (t)$ and just assume that it oscillates with frequency of $\Omega$.
We follow the discussion given in Refs.~\cite{Greene:1998nh,Peloso:2000hy,Asaka:2010kv}.
See also Appendix A in Ref.~\cite{Ema:2016hlw}.

\subsection{Quantization}
Let us start with the mode expansion of the Majorana field:
\begin{align}
	\psi (x) = \int \frac{\dd^3 k}{ (2 \pi)^{3/2} } \, e^{i \vec{k} \cdot \vec{x}} \psi_{\vec{k}} (t).
\end{align}
The mode function obeys
\begin{align}
	0 = \com{ \der_0 \gamma^0 -i\vec{k} \cdot \vec{\gamma} - m (t) } \psi_{\vec{k}} (t). 
	\label{eq:app_eom}
\end{align}
The Majorana condition, $\psi (x) = -C^{-1} \overline \psi^T (x)$,\footnote{
	Here we match the definition of the charge conjugation matrix 
	$C$ with that of Ref.~\cite{Freedman:2012zz}.
} puts the following constraints on the mode function:
\begin{align}
	\psi_{\vec{k}} (t) = -C^{-1} \overline{\psi}_{- \vec{k}}^T (t).
\end{align}
To quantize the Majorana field, we introduce the creation/annihilation operator:
\begin{align}
	\psi_{\vec{k}} (t) = \sum_s \com{
		u_{\vec{k},s} (t) \hat b_{\vec{k},s} + v_{\vec{k},s} (t) \hat b^\dag_{-\vec{k},s}
	},
\end{align}
where $v_{\vec{k},s} (t) = -C^{-1} \overline u_{-\vec{k},s}^T (t)$.
One can see that the Majorana condition is fulfilled. 
We take the following normalization of the spinor:
\begin{align}
	u_{\vec{k},s}^\dag (t) u_{\vec{k},s'} (t) = \delta_{s s'},
	\quad
	v_{\vec{k},s}^\dag (t) v_{\vec{k},s'} (t) = \delta_{s s'}, \quad u_{\vec{k},s}^\dag (t) v_{\vec{k},s'} (t) = 0.
\end{align}
The quantization condition comes from the equal time anti-commutators:
\begin{align}
	\com{ \psi (t , \vec{x}), \psi^\dag (t, \vec{y}) }_+ = \delta (\vec{x} - \vec{y}),
	\quad
	\com{ \psi (t , \vec{x}), \psi (t, \vec{y}) }_+ = 	\com{ \psi^\dag (t , \vec{x}), \psi^\dag (t, \vec{y}) }_+ = 0.
\end{align}
Together with the normalization of spinors, we obtain the following algebras for creation/annihilation operators:
\begin{align}
	\com{ \hat b_{\vec{k},s}, \hat b_{\vec{k'},s'}^\dag }_+ = \delta (\vec{k} - \vec{k'}) \delta_{s s'}, \quad
	\com{ \hat b_{\vec{k},s}, \hat b_{\vec{k'},s'} }_+ =
	\com{ \hat b_{\vec{k},s}^\dag, \hat b_{\vec{k'},s'}^\dag }_+ = 0.
\end{align}

Hereafter we adopt the Dirac representation for gamma matrices\footnote{
	\begin{align}
		\gamma_0 = \left(\begin{array}{cc}
			i & 0 \\
			0 & -i
			\end{array}\right),
			\quad
		\vec{\gamma} = \left(\begin{array}{cc}
			0 & i\vec{\sigma} \\
			-i\vec{\sigma} & 0
			\end{array}\right), 
			\quad
		\gamma_5 = \left(\begin{array}{cc}
			0 & 1 \\
			1 & 0
			\end{array}\right),
			\quad
		C = -i\gamma^0 \gamma^2.
	\end{align}
}
and expand the spinor in terms of helicity: 
\begin{align}
	u_{\vec{k},h} (t) = 
	\left(\begin{array}{c}
	u^+_{\vec{k},h} (t) \\ 
	u^-_{\vec{k},h} (t)\end{array}\right) 
	\otimes \xi_{\vec{k},h}, 
	\quad
	v_{\vec{k},h} (t) = 
	\left(\begin{array}{c}
	- u^-_{- \vec{k},h} (t) \\ 
	u^+_{- \vec{k},h} (t)\end{array}\right)^*
	\otimes \xi_{-\vec{k},h}'.
\end{align}
Here $\xi_{\vec{k},h}$ is the normalized eigenvector of helicity $h$,
satisfying $(\vec{\sigma} \cdot \hat{\vec{k}} ) \xi_{\vec{k},h} = h \xi_{\vec{k},h}$.
$\hat{\vec{k}} \equiv \vec{k} / k$ is a unit vector.
We have also defined $\xi'_{\vec{k},h} \equiv - i \sigma^2 \xi_{\vec{k},h}^*$,
which satisfies $(\vec{\sigma} \cdot \hat{\vec{k}} ) \xi'_{\vec{k},h} = - h \xi'_{\vec{k},h}$.
The normalization condition requires
\begin{align}
	\abs{u^+_{\vec{k},h}}^2 + \abs{u^-_{\vec{k},h}}^2 = 1.
	\label{eq:app_norm_pn}
\end{align}
Let us write down the equation of motion, Eq.~\eqref{eq:app_eom}, by using $u^+_{\vec{k},h}$ and $u^-_{\vec{k},h}$:
\begin{align}
	i \der_0 u^+_{\vec{k},h} + h k u^-_{\vec{k},h} &= m (t) u^+_{\vec{k},h}, 
	\label{eq:app_pos}\\
	i \der_0 u^-_{\vec{k},h} + h k u^+_{\vec{k},h} &= - m (t) u^-_{\vec{k},h}.
	\label{eq:app_neg}
\end{align}
The equations of motion for $u^{\pm}_{\vec{k},h}$ is reduced to the following linear one:
\begin{align}
	0 = \ddot u^+_{\vec{k},h} (t) + \com{ \omega_{\vec{k}}^2 (t) + i \dot m (t)  } u^+_{\vec{k},h} (t)
	=: \ddot u^+_{\vec{k},h} (t) + \widetilde \omega_{\vec{k}}^2 (t) u^+_{\vec{k},h} (t),
	\label{eq:app_eom_pos}
\end{align}
where $\omega_{\vec{k}}^2 (t) = m(t)^2 + k^2$.
From the solution of this equation, we can formally obtain $u^-_{\vec{k},h}$ via Eq.~\eqref{eq:app_neg}.
We are mainly interested in the vacuum initial condition,
which is annihilated by $b_{\vec{k},h}$.
\begin{align}
	u^+_{\vec{k},h} (t \to 0) = \sqrt{\frac{\omega_{\vec{k}} (0) + m(0)}{2  \omega_{\vec{k}} (0)}}, \quad
	\dot u^+_{\vec{k},h} (t \to 0) = - i  \omega_k (0) u^+_{\vec{k},h} (t \to 0).
	\label{eq:initial_+}
\end{align}
From Eqs.~\eqref{eq:app_pos} and \eqref{eq:app_neg}, one gets the initial condition for $u^-_{\vec{k},h}$
\begin{align}
	u^-_{\vec{k},h} (t \to 0) 
		= - h \sqrt{ \frac{\omega_{\vec{k}} (0) - m (0)}{2 \omega_{\vec{k}} (0)} }, \quad
	\dot u^-_{\vec{k},h} (t \to 0) = - i \omega_{\vec{k}} (0) u^-_{\vec{k},h} (t \to 0).
	\label{eq:initial_-}
\end{align}
Note that those are consistent with the normalization condition, Eq.~\eqref{eq:app_norm_pn}.

\subsection{Particle production}
Now we are in a position to discuss particle production. 
Let us start with evolution of the energy density.
Since we are interested in particle production from the vacuum of Majorana fermion,
we take the state which is annihilated by $\hat b_{\vec{k},h} \ket{0} = 0$.
The expectation value of the Hamiltonian density with respect to vacuum is given by
\begin{align}
	\vev{ \mathcal H (t) } &= \frac{1}{2} \vev{ \psi^\dag i \der_0 \psi } \nonumber \\[.5em]
	&= \int \frac{\dd^3 k}{ (2 \pi)^3 }\, \frac{1}{2} \sum_h
	\com{
		m (t) \prn{ \abs{u^-_{\vec{k},h} (t)}^2 - \abs{u^+_{\vec{k},h} (t)}^2}
		+ 2 h k \Re \prn{ u^+_{\vec{k},h} (t) u^{-^*}_{\vec{k},h} (t) }
	},
	\label{eq:app_energy_dens}
\end{align}
where we have used the equation of motion for $\psi$ in the first equality,
and the bra-ket stands for the expectation value with respect to vacuum.
One can check that the energy density only contains vacuum fluctuations 
for the initial condition given by Eqs.~\eqref{eq:initial_+} and \eqref{eq:initial_-}:
\begin{align}
	\vev{\mathcal H (0)} = \int \frac{\dd^3 k}{ (2 \pi)^3 }\, \sum_h \prn{ -  \frac{\omega_{\vec{k}} (0)}{2} }
	= 2 \times \int \frac{\dd^3 k}{ (2 \pi)^3 }\, \prn{- \frac{\omega_{\vec{k}} (0)}{2}}.
	\label{eq:app_energy_dens_ini}
\end{align}
It is noticeable that the integrand in the square bracket of Eq.~\eqref{eq:app_energy_dens} does not
depend on helicity $h$ under the initial condition, Eqs.~\eqref{eq:initial_+} and \eqref{eq:initial_-},
and thus, the summation over helicity becomes trivial.
This is because we have $u^+_{\vec{k},h} (t) = u^+_{\vec{k},-h}$ and $u^-_{\vec{k},h} (t) = - u^-_{\vec{k},-h} (t)$,
as one can see from Eqs.~\eqref{eq:app_eom_pos}, \eqref{eq:app_pos} and \eqref{eq:initial_+}.
Those equations, Eqs.~\eqref{eq:app_energy_dens} and \eqref{eq:app_energy_dens_ini}, 
motivate us to define the following number density:
\begin{align}
	n_\psi (t) = 2 \times \int \frac{\dd^3 k}{(2\pi)^3} \, f_\psi (\vec{k};t),
	\label{eq:app_number}
\end{align}
where the phase space density is given by
\begin{align}
	f_\psi (\vec{k}; t) 
	&\equiv 
	\frac{1}{2 \omega (t)} 	\com{
		m (t) \prn{ \abs{u^-_{\vec{k}} (t)}^2 - \abs{u^+_{\vec{k}} (t)}^2}
		+ 2 h k \Re \prn{ u^+_{\vec{k}} (t) u^{-^*}_{\vec{k}} (t) }
	} + \frac{1}{2} \nonumber \\[.5em]
	& = \frac{1}{2 \omega (t)} \com{ m(t) + 2 \Im \prn{ u^{+^*}_{\vec{k}} (t) \der_0 u^+_{\vec{k}} (t) }} + \frac{1}{2}.
	\label{eq:app_phase_space}
\end{align}
We have omitted the helicity subscript, $h$, in the wave function, $u^\pm$,
because $f_\psi$ does not depend on helicity.
The factor two in Eq.~\eqref{eq:app_number} represents the degrees of freedom for the Majorana fermion.

We assume the following form of the solution to Eq.~\eqref{eq:app_eom_pos}:
\begin{align}
	u^+_{\vec{k},h} (t) = 
	\frac{A_{k,h} (t)}{\sqrt{2 \widetilde \omega_{\vec{k}} (t)}} 
	e^{- i \int^t \dd \tau \widetilde \omega_{\vec{k}} (\tau)}
	+ \frac{B_{k,h} (t)}{\sqrt{2 \widetilde \omega_{\vec{k}}(t)}} e^{ i \int^t \dd \tau \widetilde \omega_{\vec{k}} (\tau)},
\end{align}
where
\begin{align}
	\dot A_k (t) = \frac{\dot{\widetilde \omega}_{\vec{k}} (t)}{2 \widetilde \omega_{\vec{k}} (t)} 
	e^{2 i \int^t \dd \tau \widetilde \omega_{\vec {k}} (\tau) } B_k (t), 
	\quad
	\dot B_k (t) = \frac{\dot{\widetilde \omega}_{\vec{k}} (t)}{2 \widetilde \omega_{\vec{k}} (t)} 
	e^{- 2 i \int^t \dd \tau \widetilde \omega_{\vec {k}} (\tau) } A_k (t).
	\label{eq:app_aandb}
\end{align}
Initial values of $A_{k,h}$ and $B_{k,h}$ are obtained as
\begin{align}
	A_k (t \to 0) = \sqrt{\omega_{\vec{k}}(0) + m(0)}, \quad B (t \to 0) = 0.
\end{align}
Let us estimate the growth rate of $f_\psi$ at the very beginning, $f_\psi \ll 1$.
At that time, we expect $A_{\vec{k}} \simeq \sqrt{\omega_{\vec{k}} + m}$ and $B_{\vec{k}} \simeq 0$ at the leading order.
We take into account the growth of $B_{\vec{k}}$ perturbatively by using Eq.~\eqref{eq:app_aandb}.
Under this assumption, one can easily obtain
\begin{align}
	B_{\vec{k}} (t) 
	&\simeq \int^t_0 \dd t' \frac{m \dot m + i \ddot m /2}{2 \widetilde\omega_{\vec{k}}^2}  
	A_{\vec{k}} (0)
	e^{ - 2 i \int^{t'} \dd \tau \widetilde \omega_{\vec{k}} (\tau)} \nonumber \\
	&\simeq
	 - i
	 A_{\vec{k}} (0)
	 \int^t_0 \dd t' \, 
	 m(t')\,
	 e^{ - 2 i \omega_{\vec{k}} t' }.
\end{align}
In the second similarity, we have performed integration by parts and assumed $k^2 \gg m^2$.
For given time $t$, the integration cancels out due to oscillations of the phase
except for $\Omega - \Delta \Omega \lesssim 2 \omega_{\vec{k}} \lesssim \Omega + \Delta \Omega$
with $\Delta \Omega \sim 1/ t$.
Within this frequency range, the phase of $m (t)$ and $e^{- 2 i \omega_{\vec{k}} t}$ cancels 
and $B_{\vec{k}}$ grows linearly with time:
\begin{align}
	B_{\vec{k}} (t) \simeq - \frac{i}{2} A_{\vec{k}} (0) \widetilde m t 
	~~ \text{for}~~ \Omega - \frac{1}{t} \lesssim 2 \omega_{\vec{k}} \lesssim \Omega + \frac{1}{t}.
\end{align}
Here and hereafter $\widetilde m$ stands for the amplitude of oscillating $m (t)$.
Similar arguments lead to
\begin{align}
	A_{\vec{k}} (t) \simeq A_{\vec{k}} (0) - i B_{\vec{k}}' (0) \widetilde m \frac{t^2}{4} 
	\simeq A_{\vec{k}} (0) \prn{1 - \frac{\widetilde m^2 t^2}{8}}
	~~ \text{for}~~ \Omega - \frac{1}{t} \lesssim 2 \omega_{\vec{k}} \lesssim \Omega + \frac{1}{t}.
\end{align}

Plugging those approximated solutions into Eq.~\eqref{eq:app_phase_space}, we get
\begin{align}
	f_\psi (\vec{k}; t) \simeq \frac{\widetilde m^2 t^2}{4}
	~~ \text{for}~~ \Omega - \frac{1}{t} \lesssim 2 \omega_{\vec{k}} \lesssim \Omega + \frac{1}{t}.
\end{align}
This expression is valid as long as $f_\psi \ll 1$, namely $q \Omega t \lesssim1$ with the resonance parameter
being $q \equiv \widetilde m^2/ \Omega^2 \ll 1$.
Finally, performing the phase space integral in Eq.~\eqref{eq:app_number},
we obtain the master equation for Majorana fermion production due to its oscillating mass term $m(t)$:
\begin{align}
	n_\psi (t) \simeq \frac{\mathcal{C}}{16 \pi} \Omega^2 \widetilde m^2 t.
	\label{eq:app_master_number}
\end{align}
Here we have introduced an order one factor $\mathcal{C}$, which depends on how $m (t)$ oscillates.
For instance, in the case of $m (t) \propto \cos\, ( \Omega t )$, we have $\mathcal{C}=1$.

Now let us assume 
that the oscillation of $m(t)$ is caused by oscillating (canonical) scalar field $\phi \simeq\phi_{\rm amp}\cos(m_\phi t)$.
The result (\ref{eq:app_master_number}) may be interpreted as the decay or annihilation of $\phi$ into $\psi$, depending on
how $m (t)$ oscillates with time.
Suppose that $m (t) \propto \phi^n (t)$, which involves $\Omega = j m_\phi$ with $j=n,n-2,n-4,\dots$.
One can see that allowed processes depend on the parity of $n$.
To avoid unnecessary complications, let us consider the case of $n = 1$ and $n= 2$ as an illustration.
In this case, the relevant frequency is $\Omega = m_\phi (2 m_\phi)$ for $n = 1 (2)$.
For $\Omega = m_\phi (2 m_\phi)$, the decay (annihilation) rate is estimated as
\begin{align}
	\Gamma(n \times \phi \to\psi\psi) \sim \frac{n n_\psi}{2n_\phi t}\sim \frac{n^3 \mathcal{C}}{32 \pi} \frac{\widetilde m^2}{\phi_{\rm amp}^2}  m_\phi
	\sim \frac{n^5 \mathcal{C}}{32 \pi} \frac{q m_\phi^3}{\phi_{\rm amp}^2}.
	\label{gamma_general}
\end{align}
Here we have substituted $n_\phi\sim m_\phi \phi_{\rm amp}^2 $.
Actually if $m (t) \propto \phi\propto\cos\, (m_\phi t )$, this coincides with the perturbative decay rate of $\phi$ into two $\psi$ particles
calculated in the standard method in quantum field theory.
In this case, the particle production can be reasonably interpreted as the decay of $\phi$.
On the other hand, if $m(t) \propto \phi^2$, this may be rather regarded as the annihilation of $\phi$ into $\psi$ particles,
with an annihilation rate of $\Gamma(\phi\phi\to\psi\psi) \sim n_\phi \sigma v$ with $\sigma v \sim (\widetilde m/\phi_{\rm amp}^2)^2$.
In the main text, we use the formula (\ref{gamma_general}) for the gravitino production rate rather than (\ref{eq:app_master_number}),
since reinterpreting the gravitino production as if it is caused by the inflaton decay may help readers understand the underlying physics.

Here are some comments. 
The calculations here are assuming that the background field (inflaton) is spatially homogeneous.
This assumption is not always valid particularly if it is steeper than the quadratic one~\cite{Greene:1997fu,Lozanov:2016hid}.\footnote{
	On the other hand, if the potential is flatter than the quadratic one, as required to be consistent with Planck observation for large field value,
	metastable localized objects called oscillons or I-balls may be produced~\cite{Bogolyubsky:1976yu,Gleiser:1993pt,Copeland:1995fq,McDonald:2001iv,Kasuya:2002zs,Amin:2010dc}.
	Once oscillons are formed, they behave as pressureless  matter.
	However, the necessary condition of its production and its phenomenological consequences are yet unclear~\cite{Gleiser:2008ty,Gleiser:2009ys,Hertzberg:2010yz,Amin:2011hj,Gleiser:2011xj,Mukaida:2014oza,Takeda:2014qma}.
We postpone a detailed investigation in this case.
}
For the quartic inflaton potential $V \propto \phi^4$, for example, inflaton fluctuations with momenta of its effective mass scale
develop due to the parametric resonance of the inflaton fluctuation itself
and the initially homogeneous configuration may mostly become semi-relativistic waves.
However, it does not much affect the estimate of resulting fermion abundance, since the equation of state of the universe remains the same
no matter how the parametric resonance is efficient~\cite{Lozanov:2016hid} and the fermion production rate is also 
roughly the same even if the production is caused by the decay/annihilation of the inflaton quanta, as explicitly shown above.
Note also that in the most relevant case, \textit{i.e.,} the transverse gravitino production in the single-superfield inflaton case, 
the abundance is dominated by just the first few oscillations after inflation during which the coherence of the inflaton is maintained.
Therefore, the gravitino abundance shown in Fig.~\ref{fig:Y} is not much affected by this subtlety.
For a higher power $V \propto \phi^n$ with $n>4$, we need much more care on the particle production rate,
since in such a case the inflaton fluctuation develops and the equation of state may approach to the radiation-dominated one~\cite{Lozanov:2016hid},
which significantly modifies the naive estimate obtained by the assumption that the background is dominated by the homogeneous inflaton condensation.
It also means that inflaton quanta may be highly relativistic so that the production rate may be suppressed by the Lorentz factor.
A careful investigation of particle production rate in this situation is beyond the scope of this paper.

\section{Gravitino production in small-field inflation models} \label{sec:small-field}

In this section we briefly comment on the gravitino production in small-field inflation models,
such as new-inflation or hybrid inflation models.
The known results of Refs.~\cite{Endo:2006tf,Endo:2007sz} are reproduced in our framework.

\subsection{Single-superfield model}

Let us consider a single-field new inflation model~\cite{Izawa:1996dv} as an example of small-field inflation models.
The K\"ahler and superpotential are assumed to be
\begin{align}
	&K = |\phi|^2 + |z|^2 - \frac{|z|^4}{\Lambda^2}, \\
	&W=\phi\left( M^2 - \frac{\lambda\phi^n}{n+1} \right) + \mu^2 z.
\end{align}
In this model, at the potential minimum $\left< \phi\right> = (M^2/\lambda)^{1/n}$, the superpotential takes a finite value:
\begin{align}
	W_0 = \frac{n}{n+1}\left< \phi\right> M^2  = m^0_{3/2} \Mpl^2.
\end{align}
After a few Hubble time after inflation ends, the oscillation amplitude of the inflaton becomes much smaller than $\left<\phi\right>$.
Expanding $\phi = \left<\phi\right>+\delta\phi$, K\"ahler and superpotentials are expressed as
\begin{align}
	&K = \left<\phi\right>(\delta\phi + \delta\phi^\dagger) + |\delta\phi|^2 + |z|^2-\frac{|z|^4}{\Lambda^2},\\
	&W \simeq  \frac{1}{2}m_\phi(\delta\phi)^2 + \mu^2 z + W_0 - m^0_{3/2}\left<\phi\right>\delta\phi,
	\label{W_single_small}
\end{align}
where the inflaton mass around the potential minimum is given by $m_\phi = nM^2/\left<\phi\right>$.
This is the same as the single-field chaotic inflation model with linear term in the K\"ahler potential studied in Sec.~\ref{sec:comment}
after the identification $c \to \left<\phi\right>$, except for the small linear $\delta\phi$ term in the superpotential.\footnote{
	The linear term in $\delta\phi$ appears because $W \simeq W_0+ W_\phi\delta\phi$ and $W_\phi = -K_\phi W/\Mpl^2$ 
	at the potential minimum.
}
Although the linear $\delta\phi$ term in the superpotential can induce the inflaton decay, the dominant contribution comes from the
inflaton-induced $z$ oscillation as studied in Sec.~\ref{sec:comment}.
The inflaton decay rate into the longitudinal gravitino pair and the resultant gravitino abundance is consistent with~\cite{Endo:2006tf,Endo:2007sz}.

\subsection{Multi-superfield model}

Next, we consider multi-field new inflation model~\cite{Asaka:1999jb,Senoguz:2004ky}:
\begin{align}
	&K = |\phi|^2 + |X|^2 + |z|^2 - \frac{|z|^4}{\Lambda^2}, \\
	&W= X(M^2-\lambda\phi^n) + \mu^2 z + W_0.
\end{align}
The potential minimum is $\left<X\right>\simeq 0$ and $\left<\phi\right>\simeq (M^2/\lambda)^{1/n}$.
Expanding the field as $\phi = \left<\phi\right> + \delta\phi$, the K\"ahler and superpotentials can be written as
\begin{align}
	&K = \left<\phi\right>(\delta\phi + \delta\phi^\dagger) + |\delta\phi|^2 + |X|^2 + |z|^2-\frac{|z|^4}{\Lambda^2},\\
	&W \simeq m_\phi X \delta \phi + \mu^2 z + W_0,
\end{align}
where the inflaton mass around the potential minimum is given by $m_\phi = nM^2/\left<\phi\right>$.
Thus the gravitino production in this theory is also the same as that in the chaotic inflation model with linear term in the K\"ahler potential
after the identification $c \to \left<\phi\right>$ in Sec.~\ref{sec:comment}.
Although the model of Sec.~\ref{sec:comment} does not have a stabilizer $X$, the discussion is almost parallel,
considering that there is a mixing term $\sim X^* z$ in the scalar potential and also $\phi$ and $X$ are maximally mixed with each other
at least for $H\lesssim m_{3/2}$.
As a result, the inflaton decay rate into the longitudinal gravitino pair and its abundance is consistent with~\cite{Endo:2006tf,Endo:2007sz}.

\section{Multi-field scalar dynamics}  \label{sec:dyn}

Let us consider the potential of real scalar $\phi_1$ and $\phi_2$:
\begin{align}
	V = \frac{1}{2}(\phi_1~ \phi_2) \mathcal M^2 \begin{pmatrix}
  	\phi_1 \\
	\phi_2
  \end{pmatrix},~~~
  \mathcal M^2 =
  \begin{pmatrix}
  	m_1^2 & m_{12}^2 \\
	m_{12}^2 & m_2^2
  \end{pmatrix} .
\end{align}
We want to estimate how large is the induced oscillation amplitude of $\phi_2$ starting with the initial condition $(\phi_1,\phi_2)=(\phi_{i},0)$.
We assume $|m_1m_2| > m_{12}^2$ so that the potential is positive definite.

To solve the equation of motion, It is convenient to move to the mass eigenstate basis:
\begin{align}
	\begin{pmatrix}
  	\phi_1' \\
	\phi_2'
  \end{pmatrix}
  =
  \begin{pmatrix}
  	c_\theta & s_\theta \\
	-s_\theta & c_\theta
  \end{pmatrix} 
  \begin{pmatrix}
  	\phi_1 \\
	\phi_2
  \end{pmatrix},
\end{align}
where $c_\theta \equiv \cos\theta$ and $s_\theta \equiv \sin\theta$ with $-\pi/4 < \theta \leq \pi/4$.
They are given by
\begin{equation}
	c_\theta^2 = \frac{1}{2}\left(1+\sqrt{1- \frac{4m_{12}^4}{4m_{12}^4+(m_1^2-m_2^2)^2} }\right),~~
	s_\theta^2 =  \frac{1}{2}\left(1-\sqrt{1- \frac{4m_{12}^4}{4m_{12}^4+(m_1^2-m_2^2)^2} }\right),
\end{equation}
with $\theta \geq 0$ for $(m_1^2-m_2^2)/m_{12}^2 > 0$ and $\theta < 0$ for $(m_1^2-m_2^2)/m_{12}^2 < 0$.
Hence the mixing angle is approximated as
\begin{equation}
	\theta \simeq \begin{cases}
		m_{12}^2/(m_1^2-m_2^2) & {\rm ~~for~~} |2m_{12}^2| \ll |m_1^2-m_2^2|,\\
		\pi/4       & {\rm ~~for~~} |2m_{12}^2| \gg |m_1^2-m_2^2|.
	\end{cases}
\end{equation}
In the mass eigenstate basis, the potential looks like
\begin{align}
	V = \frac{1}{2}(\phi_1'~ \phi_2') \mathcal M^{'2} \begin{pmatrix}
  	\phi_1' \\
	\phi_2'
  \end{pmatrix},~~~
  \mathcal M^{'2} =
  \begin{pmatrix}
  	m_1^{'2} & 0 \\
	0  & m_2^{'2}
  \end{pmatrix},
\end{align}
where
\begin{align}
	m_1^{'2} &= \frac{1}{2}\left( m_1^2+m_2^2 + \frac{|m_1^2-m_2^2|}{m_1^2-m_2^2}\sqrt{ (m_1^2-m_2^2)^2+4m_{12}^4 } \right),\\
	m_2^{'2} &= \frac{1}{2}\left( m_1^2+m_2^2 - \frac{|m_1^2-m_2^2|}{m_1^2-m_2^2}\sqrt{ (m_1^2-m_2^2)^2+4m_{12}^4 } \right).
\end{align}
In this basis, the equation of motion of each field can be easily solved as
\begin{align}
	\phi_1'(t) &= c_\theta \phi_i \left(\frac{a_i}{a(t)}\right)^{3/2}\cos(m_1' t),\\
	\phi_2'(t) &= -s_\theta \phi_i \left(\frac{a_i}{a(t)}\right)^{3/2}\cos(m_2' t),
\end{align}
under the initial condition $(\phi_1,\phi_2)=(\phi_{i},0)$.
Then we obtain
\begin{align}
	\phi_1(t) &= \phi_i  \left(\frac{a_i}{a(t)}\right)^{3/2} \left[ c_\theta^2 \cos(m_1' t) + s_\theta^2\cos(m_2' t) \right],\\
	\phi_2(t) &=- \phi_i  \left(\frac{a_i}{a(t)}\right)^{3/2}\sin(2\theta)\sin\left(\frac{(m_1'+m_2')t}{2} \right)\sin\left(\frac{(m_1'-m_2')t}{2} \right).
\end{align}

Therefore, if $m_1'$ and $m_2'$ are not close to each other, we can just regard $\phi_2(t) \sim \sin(2\theta) \phi_i$ as an ``induced'' oscillation amplitude
after a few oscillation of each field.
On the other hand, if $m_1' \simeq m_2'$, or $m_1^2 \simeq m_2^2 \gg |m_{12}^2|$ in the original basis,
it takes long time to develop large amplitude of $\phi_2$.
In the degenerate limit $m_1=m_2$, we obtain $m_1' - m_2' \simeq m_{12}^2/m_1$.
In this case, we have
\begin{align}
	\phi_2(t) \simeq - \phi_i  \left(\frac{a_i}{a(t)}\right)^{3/2}\sin(m_1t)\, \frac{m_{12}^2}{2m_1}t ~~~{\rm for}~~~t \lesssim \frac{2m_1}{m_{12}^2}.
\end{align}
Hence after the time $t \sim 2m_1/m_{12}^2$, $\phi_1$ and $\phi_2$ may be regarded as maximally mixed with each other.

\section{Calculation of mass eigenvalues} \label{sec:mass}
In this appendix, we estimate the mass eigenvalues of the system of the longitudinal gravitino and matter fermions discussed in Secs.~\ref{sec:long_single} and \ref{sec:long_multi}.
\subsection{Single-superfield model}
In this subsection, we consider the model in Sec.~\ref{sec:single}, which does not utilize the stabilizer field.
The full mass matrix $\mathcal{M}$ is given by Eq.~\eqref{M_2field}.
We regard the first and second row/column (thus corresponding subscripts or superscripts 1 and 2) correspond to the inflaton(ino) $\phi$ and the Polonyi(no) field $z$, respectively.
The trace and determinant of the mass matrix are 
\begin{align}
\tr\, \mathcal{M} =& \widehat{m}_{3/2} + m_f -\alpha_1^2 \dot{\theta}_1 - \alpha_2^2 \dot{\theta}_2 \nonumber \\
\simeq &  m_{\phi}+2\alpha_1^2 \widehat{m}_{3/2}^1 + (\alpha_{1}^2 - \alpha_{2}^2) m_{3/2}, \\
\det  \, \mathcal{M} = & \alpha_1^2 \alpha_2^2 \sin^2 (\theta_1 - \theta_2) (\widehat{m}_{3/2} - m_f^c )^2 -\alpha_2 \dot{\alpha}_2 (\widehat{m}_{3/2} - m_f^c ) \sin (2 (\theta_1 - \theta_2)) \nonumber \\
& -(\dot{\alpha}_1^2 +\dot{\alpha}_2^2)\sin^2 (\theta_1 -\theta_2) + \alpha_2^2 (\widehat{m}_{3/2} - m_f^c )(\dot{\theta}_2 - \dot{\theta}_1) + (\widehat{m}_{3/2} -\dot{\theta}_1 ) ( m_f^c -\dot{\theta}_2 ) \nonumber \\
\simeq &  - \alpha_2^2 m_{\phi} m_{3/2} +\alpha_1^2 \alpha_2^2 \sin^2(\theta_1 -\theta_2) (\widehat{m}_{3/2}^1 )^2  -\alpha_1^2 \alpha_2^2 \left( \widehat{m}_{3/2}^1  + m_{3/2} \right)^2 ,
\end{align}
where $m_f^c \equiv m_f + \alpha_2^2 \dot{\theta}_1 + \alpha_1^2 \dot{\theta}_2 \simeq -\alpha_2^2 (m_{3/2} + \widehat{m}_{3/2}^1)$ is introduced as a useful combination, which sits on the (2, 2) component of the inner parenthesis in Eq.~\eqref{M_2field}, and it satisfies $\widehat{m}_{3/2}-m_f^c \simeq \widehat{m}_{3/2}^1$.
In the above estimation, we have used $\dot{\theta}_{\phi} \simeq  - m_{\phi} - m_{3/2} - \widehat{m}_{3/2}^{1}$ and $\dot{\theta}_{z} \simeq  0$ (\textit{cf.}~Eq.~\eqref{theta_dot_i}).
The quantities $\widehat{m}_{3/2}^i$ can be estimated using Eq.~\eqref{mhati_def} as $\widehat{m}_{3/2}^1 = (m_{3/2}+3 H \sin 2 \theta_1 - 3 m_{3/2} \cos 2 \theta_1 )/2$ and $\widehat{m}_{3/2}^2 \simeq - m_{3/2}$.
The time-derivative of the SUSY breaking fractions $\dot{\alpha}_i$ have been neglected because $\alpha_1 \dot{\alpha}_1 = - \alpha_2 \dot{\alpha}_2 = \mathcal{O}(\min \left[ (m_{3/2}^0)^2/H , H^2 / (m_{3/2}^0 \right ]) \lesssim \mathcal{O}(m_{3/2}^0)$.
Note that the trace contains a violently oscillating term $\widehat{m}_{3/2}^1$ while the determinant does not have a term of order $m_{\phi} \widehat{m}_{3/2}^1$ (\textit{cf.}~\ref{ftn:m32hat_vs_m32}).

Since $(\tr \, \mathcal{M})^2 \gg \det \, \mathcal{M}$, the mass eigenvalues of $\mathcal{M}$ are given by
\begin{align}
\frac{1}{2}\left( \tr \, \mathcal{M} \pm \sqrt{(\tr \, \mathcal{M})^2 - 4 \det \, \mathcal{M}} \right)= 
\begin{cases}
\tr\, \mathcal{M}    & \text{for the heavy state}, \\
\det \, \mathcal{M} / \tr\, \mathcal{M} & \text{for the light state}.
\end{cases}
\end{align}
Thus, we obtain the eigenvalues $(m_{\text{heavy}}, m_{\text{light}})\simeq (m_{\phi}+2\alpha_1^2 \widehat{m}_{3/2}^1, - \alpha_2^2 m_{3/2})$.
Keeping the first order of $\delta z$, we find the Polonyi-dependent part of the light mass eigenvalue $-\dot{\theta}_2$.  That is, $m_{\text{light}}\simeq - \alpha_2^2 m_{3/2} + \frac{m_z^2}{\sqrt{3} m_{3/2}^0 \Mpl}\delta z$, which is used in eqs.~\eqref{mlight_early} and \eqref{mlight}.

\subsection{Multi-superfield model}
Here, we evaluate the mass matrix $\mathcal{M}$ in Eq.~\eqref{mass_multi} in the case with the stabilizer field in Sec.~\ref{sec:multi}.
The matter fermion part of the mass matrix is
\begin{align}
\widehat{m}_f \simeq m_f \simeq \text{diag} ( m_{\phi}, - m_{\phi}, 0 ),
\end{align}
in the ``light-cone'' basis $(\Phi_+, \Phi_-, z)$ defined in Eq.~\eqref{light-cone_basis}.

The trace is 
\begin{align}
\tr \mathcal{M} = (1-2 \alpha_{z}^2) m_{3/2}+ 2 \left( \alpha_{+}^2 \widehat{m}_{3/2}^+ +  \alpha_{-}^2 \widehat{m}_{3/2}^- \right),
\end{align}
where the term of order $m_{\phi}$ cancels out because of the opposite phases.
The full expression of the determinant is long and complicated.
Since we expect there are two states with their mass eigenvalues approximately equal to $\pm m_{\phi}$, we extract the terms proportional to $m_{\phi}^2$,
\begin{align}
\det  \mathcal{M} = s_2^2  m_{3/2} m_{\phi}^2 + \dots = \alpha_z^2  m_{3/2} m_{\phi}^2 + \dots.
\end{align}
Note that the lightest eigenvalue is not of order $\widehat{m}_{3/2}$ but of order $m_{3/2}$ (\textit{cf.}~\ref{ftn:m32hat_vs_m32}).
In this calculation, we have used $\dot{\theta}_{\pm} \simeq  \mp m_{\phi} - m_{3/2} - \widehat{m}_{3/2}^{\pm}$ with $\widehat{m}_{3/2}^{\pm} \simeq (m_{3/2} - 3 m_{3/2} \cos 2 \theta_1 \pm 3 H  \sin 2 \theta_1 )/2$. 
Finally, if we call the three mass eigenvalues $m_1, m_2$, and $m_3$, 
we have a relation
\begin{align}
m_1 m_2 + m_2 m_3 + m_3 m_1 =& \mathcal{M}_{11}\mathcal{M}_{22}+\mathcal{M}_{22}\mathcal{M}_{33}+\mathcal{M}_{33}\mathcal{M}_{11}-\mathcal{M}_{12}^2 -\mathcal{M}_{23}^2 - \mathcal{M}_{31}^2 \nonumber \\
=& - m_{\phi}^2 - 2( \alpha_{+}^2 \widehat{m}_{3/2}^{+} - \alpha_{-}^2 \widehat{m}_{3/2}^{-} + \mathcal{O}(m_{3/2})) m_{\phi} + \dots,
\end{align}
where we write only the largest term, which is proportional to $m_{\phi}^2$, and the subdominant but oscillating term linear in $m_{\phi}$.
Thus, we conclude that there are three states with their mass eigenvalues $(m_{\text{heavy}}^{+}, m_{\text{heavy}}^{-}, m_{\text{light}}) \simeq (m_{\phi}+2\alpha_{+}^2 \widehat{m}_{3/2}^{+}, - m_{\phi}+2\alpha_{-}^2 \widehat{m}_{3/2}^{-}, - \alpha_z^2 m_{3/2})$.
Repeating the same analysis without neglecting $\delta z$, we find the $z$-dependent part of the light eigenvalue, $m_{\text{light}} \simeq - \alpha_z^2 m_{3/2} +   \frac{m_z^2}{\sqrt{3} m_{3/2}^0 \Mpl}\delta z$.

\small
\bibliography{ref}


\end{document}